\documentclass[12pt]{article}
\usepackage{epsfig,amssymb,amsmath}
\usepackage{jheppub}
 
\usepackage{esint} 
\usepackage{breqn}

\def \tr {\mathop{\rm tr}\nolimits}

\def \e  {\mathop{\rm e}\nolimits}

\newcommand\lr[1]{{\left({#1}\right)}}

\newcommand \vev [1] {\langle{#1}\rangle}

\newcommand \ket [1] {|{#1}\rangle}
\newcommand \bra [1] {\langle {#1}|}
\newcommand\re[1]{(\ref{#1})}
\def \qqquad {\qquad\quad}
\def \qqqquad {\qquad\qquad}

\newcommand{\ft}[2]{{\textstyle\frac{#1}{#2}}}

\def\numberbysection{\@addtoreset{equation}{section}
                     \def\theequation{\thesection.\arabic{equation}}}

\preprint{\small  \parbox[t]{25mm}{IPhT-T16/087}}

\title{\Large Instanton corrections to twist-two operators 
}

\author[a]{Luis F. Alday}
\author[b]{and Gregory P. Korchemsky}

\affiliation[a]{Mathematical Institute, University of Oxford,  Andrew Wiles Building, Radcliffe Observatory Quarter, Woodstock Road, Oxford, OX2 6GG, UK}

\affiliation[b]{Institut de Physique Th\'eorique\footnote{Unit\'e Mixte de Recherche 3681 du CNRS}, Universit\'e Paris Saclay, CNRS, CEA, F-91191 Gif-sur-Yvette}

\abstract{We present the calculation of the leading instanton contribution to the scaling dimensions of twist-two operators with arbitrary
spin and to their structure constants in the OPE of two half-BPS operators in $\mathcal N=4$ SYM. For spin-two operators we verify that, 
in agreement with $\mathcal N=4$ superconformal Ward identities, the obtained expressions 
coincide with those for the Konishi operator. For operators with high spin we find that the leading instanton correction 
vanishes. This arises as the result of a rather involved calculation and requires a better understanding. 
}

\begin{document}

\maketitle

\section{Introduction} 

In this paper we continue the study of instanton corrections to correlation functions in maximally supersymmetric 
$\mathcal N=4$ Yang-Mills theory. Although these corrections are exponentially small in the planar limit, they
are expected to play an important role in restoring the $S-$duality of the theory. At weak coupling, the leading
instanton contribution can be found in the semiclassical approximation by neglecting quantum fluctuation of fields. In this approximation, the calculation amounts to evaluating the product of operators in the background of
instantons and integrating the resulting expression over the collective coordinates. For a review see \cite{Belitsky:2000ws,Dorey:2002ik,Bianchi:2007ft}. 

Previous studies revealed \cite{Bianchi:1999ge} that the leading instanton contribution to four-point correlation function of half-BPS operators in 
$\mathcal N=4$ SYM scales at weak coupling as $\e^{-8\pi^2/g^2}$. An OPE analysis showed, however, that this correction does not affect
twist-two operators \cite{Arutyunov:2000im}  and, therefore, does not modify the leading asymptotic behaviour of correlation functions
in the light-cone limit. This led to the conclusion \cite{Bianchi:2001cm,Kovacs:2003rt} that the leading instanton contribution to the conformal data of twist-two
operators (scaling dimensions $\Delta_S$ and OPE coefficients $C_S$) should be suppressed by a power of the coupling constant and scale as
$g^{2n} \e^{-8\pi^2/g^2}$ with some $n\ge1$. The calculation of such corrections within the conventional approach is way more
complicated as it requires going beyond the semiclassical approximation. 

In \cite{Alday:2016tll} we argued that, 
by virtue of $\mathcal N=4$ superconformal symmetry, the above mentioned instanton effects can be determined from the semiclassical
computation of two- and three-point correlation functions  for another operator in the same supermultiplet. Following this
approach, we computed the leading non-vanishing correction to the scaling dimension of the Konishi operator, $\Delta_K^{\rm (inst)} = O(g^4 \e^{-8\pi^2/g^2})$
and to its structure constant in the OPE of two half-BPS operators, $C_K^{\rm (inst)} =  O(g^2 \e^{-8\pi^2/g^2})$ (see  \cite{Alday:2016tll} for
explicit expressions).

In this paper we
extend the analysis to twist-two operators $O_S$ with arbitrary even Lorentz spin $S$. For spin zero, the operator $O_{S=0}$ coincides 
with the half-BPS operator and is protected from quantum corrections. For spin-two, the operator $O_{S=2}$ belongs to
the same supermultiplet as the Konishi operator and, therefore, has the same conformal data. For $S\ge 4$, quite surprisingly,  our calculation
yields a vanishing result for the instanton contribution. This implies that the leading instanton corrections to the conformal
data of twist-two operators  $O_S$ with $S\ge 4$ are suppressed  at least  by a power of $g^2$ as compared with those for the
Konishi operator
\begin{align}\notag\label{res0}
{}& \Delta_S^{\rm (inst)} = \delta_{S,2} \, \Delta_K^{\rm (inst)} + O(g^6 \e^{-8\pi^2/g^2})\,,
\\[2mm]
{}& C_S^{\rm (inst)}  =     \delta_{S,2} \, C_K^{\rm (inst)} + O(g^4 \e^{-8\pi^2/g^2})\,.
\end{align}
Notice that these two expressions differ by a power of the coupling constant, $\Delta_S^{\rm (inst)}/C_S^{\rm (inst)}=O(g^2)$,
whereas the leading perturbative corrections to both quantities have the same scaling in $g^2$  at weak coupling.

The paper is organized as follows. In Section~2 we define operators of twist two and discuss their relation to light-ray operators.
In Section~3 we construct the one-instanton solution to the equations of motion in $\mathcal N=4$ SYM for the $SU(2)$ gauge group.  
In Section~4 we present the calculation of correlation functions involving half-BPS and twist-two operators in the semiclassical approximation 
and discuss its generalization to the $SU(N)$ gauge group.
Section~5 contains concluding remarks. Some details of the calculation are summarized in four Appendices.
 
\section{Twist-two operators}
 
All twist-two operators in $\mathcal N=4$ SYM belong to the same supermultiplet and share the same conformal data. This allows us to restrict our consideration  to
the simplest  twist-two operator, of  the form
\begin{align}\label{tw-2-def}
O_S(x) = \tr\left[ Z \, D_+^S \,Z(x)\right] + \dots\,,
\end{align} 
where $Z(x)$ is a complex scalar field and $D_+= n^\mu D_\mu$  (with $n^2=0$) is a light-cone component of the 
covariant derivative $D_\mu =\partial_\mu + i[A_\mu, \ ]$. All fields take values in the $SU(N)$ 
algebra, e.g. $Z(x)=Z^a(x) T^a$ with the generators normalized as $\tr(T^a T^b)=\delta^{ab}/2$.
The dots on the right-hand side of \re{tw-2-def} denote a linear combination of operators with total derivatives 
of the form $\partial_+^{\ell}  \tr\left[ Z  D_+^{S-\ell} Z(x)\right]$ with $0\le \ell \le S$. The corresponding expansion coefficients 
are fixed by the condition for $O_S(x)$ to be a 
conformal primary operator and depend, in general, on the coupling constant. 
To lowest order in the coupling, they are related to those of
the Gegenbauer polynomials (see Eq.~\re{c-coef} below). 

In this paper, we compute the leading instanton corrections to correlation functions of twist-two operators \re{tw-2-def} and 
half-BPS scalar operators of the form
\begin{align}\label{1/2}
O_{\bf 20'}(x) = Y_{AB} Y_{CD} \tr[\phi^{AB} \phi^{CD}(x)]\,,
\end{align}
where the complex scalar fields  $\phi^{AB}=-\phi^{BA}$(with $A,B=1,\dots,4$) satisfy reality condition 
$\bar\phi_{AB} = \frac12 \epsilon_{ABCD}\phi^{CD}$. The auxiliary antisymmetric tensor $Y_{AB}$ is introduced to project the product of two scalar fields onto the representation $\bf 20'$ of the $SU(4)$ $R-$symmetry group.
It satisfies $\epsilon^{ABCD} Y_{AB} Y_{CD} =0$ and plays the role of the coordinate of the operator in the isotopic $SU(4)$ space.
The scalar field $Z$ entering \re{tw-2-def} is a special component of  $\phi^{AB}$ 
\begin{align}
Z=\phi^{14} = (Y_Z)_{AB} \phi^{AB} \,,
\end{align}
where $(Y_Z)_{AB}$ has the same properties as the $Y-$tensor in \re{1/2} and has the only nonvanishing components $(Y_Z)_{14}=-(Y_Z)_{41}=1/2$.

Conformal symmetry fixes the form of two- and three-point correlation functions of the operators \re{tw-2-def} and \re{1/2}
\begin{align}\notag\label{gen-exp}
{}& \vev{O_S(x) \bar O_{S'}(0)} = \delta_{SS'}  \mathcal N_S {[2(xn)]^{2S}\over (x^2)^{\Delta_S+S}}\,,
\\
{}& \vev{O_{\bf 20'}(x_1)O_{\bf 20'}(x_2)O_S(0)} = {C_S \over (x_{12}^2)^2 } \left[{2(nx_1)\over x_1^2} -{2(nx_2)\over x_2^2}\right]^S 
\lr{x_{12}^2\over x_1^2 x_2^2}^{(\Delta_S-S)/2}
\,.
\end{align}
Here the scaling dimension of twist-two operator $\Delta_S$, the normalization factor $\mathcal N_S$ and three-point coefficient function 
$C_S$ depend on the coupling constant whereas the scaling dimension of the half-BPS operator is protected from
quantum corrections.

\subsection{Light-ray operators}

To compute the correlation functions \re{gen-exp}, it is convenient to introduce a generating function for the twist-two operators \re{tw-2-def}, the so-called
light-ray operator,
\begin{align} \label{O-LC}
\mathbb O(z_1,z_2) {}&= \tr\Big[Z(nz_1) E(z_1,z_2) Z(nz_2) E(z_2,z_1)\Big]\,.
\end{align}
In distinction with \re{tw-2-def}, it is a nonlocal operator -- the two scalar fields are separated along the light-ray direction
$n^\mu$ and two light-like Wilson lines are inserted to restore gauge invariance,
\begin{align}\label{line}
E(z_1,z_2) = P\exp\lr{i \int_{z_1}^{z_2} dt\,  n^\mu A_\mu (nt)}\,,
\end{align}
with $E(z_1,z_2)E(z_2,z_1)=1$. The scalar variables $z_1$ and $z_2$ define the position of the fields along the null ray.

Making use of gauge invariance of \re{O-LC}, we can fix the gauge 
$n^\mu A_\mu(x)=0$ in which Wilson lines \re{line} reduce to $1$. Then, the expansion of \re{O-LC} in powers of $z_1$ and $z_2$ takes
the form
\begin{align}\label{Taylor}
\mathbb O(z_1,z_2) {}&=\sum_{k,n\ge 0} {z_1^k \over k!}  {z_2^n \over n!}\tr\big[\partial_+^k Z(0) \partial_+^n Z(0) \big]\,,
\end{align}
where $\partial_+ = (n\partial)$. To restore gauge invariance, it suffices to replace $\partial_+\to D_+$  in this relation. 
The local operators on the right-hand side of \re{Taylor} are not conformal primaries but, for given $S=k+n$, they can be 
expanded over the conformal  primary
operators $O_S(0)$ and their descendants $\partial_+^\ell O_{S-\ell}(0)$. 

As mentioned at the beginning of this section, the conformal operators have the following general form
\begin{align} \label{O-sum0}
O_S(0)  = \sum_{  k+n=S} c_{kn}  \tr\big[D_+^k Z(0) D_+^n Z(0) \big] \,, 
\end{align}
with the expansion coefficients $c_{kn}$ depending on the coupling constant. To lowest order in the coupling, these coefficients coincide 
(up to an overall normalization) with those of $(x_1+x_2)^S C_S^{1/2}((x_1-x_2)/(x_1+x_2))=\sum c_{kn} x_1^k x_2^n$ involving the 
Gegenbauer polynomial \cite{Makeenko:1980bh,Ohrndorf:1981qv}
\begin{align}\label{c-coef}
c_{kn} =(-1)^n {S!\over (k! n!)^2} + O(g^2)\,.
\end{align}
Note that  the sum in \re{O-sum0} vanishes for odd $S$ and the conformal operators are defined for even
nonnegative $S$.

Inverting \re{O-sum0} we can expand the light-ray operator \re{Taylor} over the conformal twist-two operators and their descendants.
In this way, we find that the operators $O_S(0)$ appear as the coefficients in the expansion of the light-ray operator 
\re{O-LC} in powers of $z_{12}\equiv z_1-z_2$ \cite{Belitsky:2005gr}
\footnote{The relation \re{OPE} 
has the following interpretation \cite{Belitsky:2005gr}. The light-ray 
operator \re{O-LC} transforms covariantly under the action of the collinear $SL(2;\mathbb{R})$ subgroup of the conformal group. Then, 
the relation \re{OPE} defines the decomposition of the tensor product of two $SL(2;\mathbb{R})$ representations
over the irreducible components. The coefficient functions $z_{12}^S$ are the lowest weights of these components.
}
\begin{align}\label{OPE}
\mathbb O(z_1,z_2) {}&=\sum_{S\ge 0} {z_{12}^S\over S!} \left[O_S(0) + \dots \right].
\end{align} 
Here the dots denote the contribution from descendant operators of the form $\partial_+^\ell O_S(0)$. This contribution is fixed by
conformal symmetry, see \cite{Braun:2003rp}.  

\subsection{From light-ray to twist-two operators}\label{sect:from}

The rationale for introducing \re{O-LC} is that finding instanton corrections to  light-ray operators proves to be simpler
as compared to that for twist-two operators. Then, having computed the correlation function 
\footnote{Such correlation
functions appeared in the study of the asymptotic behaviour of four-point correlation functions
$ \vev{O_{\bf 20'}(x_1)O_{\bf 20'}(x_2)O_{\bf 20'}(x_3)O_{\bf 20'}(x_4)}$
in the light-cone limit $x_{34}^2\to 0$, see \cite{Alday:2010zy}. } 
 \begin{align}\label{G4-lc}
G(x_1,x_2,nz_1,nz_2)  =  \vev{O_{\bf 20'}(x_1)O_{\bf 20'}(x_2)\mathbb O(z_1,z_2)}\,,
\end{align}
we can then apply \re{OPE} to obtain the three-point correlation function $ \vev{O_{\bf 20'}(x_1)O_{\bf 20'}(x_2)O_S(0)} $.

To lowest order in the coupling, we can simplify the calculation by making use of the following relation
between the operators $O_S(0)$ and $ \mathbb O(z_1,z_2)$
\begin{align}\label{double}
 O_S(0) =  \oiint {dz_1dz_2\over (2\pi i)^2} {(z_1-z_2)^S\over (z_1z_2)^{S+1}} 
 \mathbb O(z_1,z_2) \,,
\end{align}
where the integration contour in both integrals encircles the origin. Indeed, replacing $ \mathbb O(z_1,z_2)$ on the right-hand side with \re{Taylor} and
computing the residue at $z_1=0$ and $z_2=0$ we obtain \re{O-sum0} with $c_{kn}$ given by \re{c-coef}. 
Using the operator identity \re{double}
inside corrrelation functions we arrive at 
\begin{align}\label{use1}
\vev{O_S(0)O_{\bf 20'}(x_1)O_{\bf 20'}(x_2)} = \oiint {dz_1dz_2\over (2\pi i)^2} {(z_1-z_2)^S\over (z_1z_2)^{S+1}} \vev{\mathbb O(z_1,z_2) O_{\bf 20'}(x_1)O_{\bf 20'}(x_2) }\,.
\end{align}  
We would like to emphasize that relations \re{double} and \re{use1} hold to the lowest order in the coupling constant. Beyond this order, we have
to take into account $O(g^2)$ corrections to \re{c-coef}. 
 
Relation \re{use1} offers an efficient way of computing the correlation functions of twist-two operators. 
As an example, we show in Appendix~\ref{App:Born} how to use 
\re{use1} to obtain the correlation functions \re{gen-exp} in the Born approximation (see Eqs.~\re{3pt-Born} and \re{2pt-Born}).  

\section{Instantons in $\mathcal N=4$ SYM}
  
The general one-instanton solution to the equations of motion in $\mathcal N=4$ SYM with the
$SU(N)$ gauge group depends 
on $8N$ fermion collective coordinates, see \cite{Belitsky:2000ws,Dorey:2002ik,Bianchi:2007ft}. Among them $16$ modes are related to $\mathcal N=4$ superconformal symmetry. The remaining $8(N-2)$ fermion modes are not related to symmetries  in an 
obvious way and are usually called `nonexact modes'.  This makes the construction of the instanton solution
more involved. 

For the $SU(2)$ gauge group
the general one-instanton solution to the equations of motion of $\mathcal N=4$ SYM 
can be obtained by applying superconformal transformations to the special solution corresponding to vanishing
scalar and gaugino fields and  gauge field given by the celebrated BPST instanton  \cite{Belavin:1975fg}

\begin{align}\label{bare}\notag
{}& \phi^{(0),AB}= 
\lambda_{\alpha}^{(0),A}= \bar \lambda^{(0)}_{\dot \alpha,A}=0\,, \qqquad
\\[2mm]
{}& A_\mu^{(0)}(x-x_0) =  2{\eta_{\mu\nu}^a (x-x_0)^\nu T^a \over (x-x_0)^2+\rho^2} \,, 
\end{align}
where $\eta_{\mu\nu}^a$ are the 't Hooft symbols and the $SU(2)$ generators are related to Pauli matrices  $T^a=\sigma^a/2$.
 It depends on the collective coordinates $\rho$ and 
$x_0$ defining the size and the position of the instanton, respectively. 

In this section, we present explicit expressions for the one-instanton solution in $\mathcal N=4$ SYM for the $SU(2)$ gauge 
group. Then, in the next section, we explain how to generalize the expressions for the correlation functions \re{gen-exp}   to the case of the $SU(N)$ gauge group. 


The field configuration \re{bare} is annihilated by (the antichiral) half of the $\mathcal N=4$ superconformal generators.
Applying to \re{bare} the remaining (chiral) $\mathcal N=4$ superconformal transformations (see \re{trans1} in Appendix~\ref{app:A}), we obtain a solution to the $\mathcal N=4$ equations of motion that 
depends on $16$ fermionic collective coordinates, $\zeta_\alpha^A$ and $\bar\eta^{\dot\alpha A}$. The 
resulting expressions for gauge and scalar fields can be expanded in powers of fermion modes
\begin{align}\notag\label{dec0}
{}& A_\mu =A_\mu^{(0)}+A_\mu^{(4)}+\dots + A_\mu^{(16)}\,,
\\[2mm] 
{}& \phi^{AB} = \phi^{AB,(2)} + \phi^{AB,(6)} + \dots + \phi^{AB,(14)} \,,
\end{align}
where $A_\mu^{(n)}$ denotes the component of the gauge field that is homogenous in $\zeta_\alpha^A$ and $ \bar \eta^{\dot\alpha A}$ 
of degree $n$ and similar for scalars. Gaugino fields admit similar expansions (see \re{dec} in Appendix~\ref{app:B}) but we will not need them for our purposes. Explicit expressions for various components of \re{dec0} are given below.

By virtue of superconformal invariance, the action of $\mathcal N=4$ SYM 
evaluated on the instanton configuration \re{dec0} does not depend on the fermionic modes and coincides with
the one for pure Yang-Mills theory
\footnote{For the $SU(N)$ gauge group the instanton action \re{action} also depends on $8(N-2)$ nonexact fermion modes, see \cite{Belitsky:2000ws,Dorey:2002ik,Bianchi:2007ft}.}
\begin{align}\label{action}
S_{\rm inst}=\int d^4 x \, L(x) = - 2\pi i \tau\,,
\end{align}  
where 
$\tau$ is the complex coupling constant  
\begin{align}\label{tau}
\tau = {\theta\over 2\pi} + {4\pi i\over g^2}\,.
\end{align}
Notice that, due to our definition of the Lagrangian (see \re{La}  in Appendix~\ref{app:A}), the instanton solution \re{bare} and \re{dec0} does not depend on the coupling
constant.  

It is straightforward to work out the leading term of the expansion \re{dec0} by subsequently applying $\mathcal N=4$ superconformal 
transformations to \re{bare}, see \cite{Belitsky:2000ws}. The direct calculation of the subleading terms becomes very involved due to the complicated form of 
these transformations (see \re{trans1} in Appendix \ref{app:A}). There is, however, a more efficient approach to computing 
higher components in \re{dec0} which is presented in Appendix~\ref{app:B}. It makes use of the known properties of fields with respect to conformal symmetry, $R-$symmetry and gauge transformations and allows us to work out the expansion \re{dec0} recursively with little efforts.  

To present the resulting expressions for the instanton configuration \re{dec0} for the $SU(2)$ gauge group it is convenient to switch to spinor notation and
use a matrix representation for \re{bare} (see Appendix~\ref{app:A} for our conventions)
\begin{align}\label{A-mat}
(A_{\alpha\dot\alpha})_i{}^j = i A_{\mu}^{a}(x) (T^a)_i{}^j (\sigma^\mu)_{\alpha\dot\alpha} \,,
\end{align}
where the four-dimensional vector of $2\times 2$ matrices $\sigma^\mu=(1,i\boldsymbol{\sigma})$ and the $SU(2)$ generators $T^a=\sigma^a/2$ are built from Pauli matrices.
This field carries two pairs of indices, 
Lorentz indices ($\alpha,\dot\alpha=1,2$) and $SU(2)$ indices ($i,j=1,2$). 
Here we distinguish lower and upper $SU(2)$ indices and define the product of two matrices by
\begin{align}
(A_{\alpha\dot\alpha})_i{}^j (A_{\beta\dot\beta})_j{}^k \equiv (A_{\alpha\dot\alpha} A_{\beta\dot\beta})_i{}^k\,.
\end{align}
 All indices are raised and lowered
with the help of the antisymmetric tensor, e.g.
\begin{align}\label{A-sym}
(A_{\alpha\dot\alpha})_i{}^k = (A_{\alpha\dot\alpha})_{ij} \epsilon^{jk}
\,,\qqqquad
(A^{(0)}_{\alpha\dot\alpha})_{ij} =  {\epsilon_{i\alpha}x_{j\dot\alpha} + \epsilon_{j\alpha}x_{i\dot\alpha}\over x^2+\rho^2} \,,
\end{align}
where the second relation follows from \re{bare} and \re{A-mat}. The advantage of   $(A_{\alpha\dot\alpha})_{ij}$ as compared
to \re{A-mat} is that it is symmetric with respect to the $SU(2)$ indices. 

The instanton \re{bare} and \re{A-sym} is a self-dual solution to the equations of motion in pure Yang-Mills theory, 
$F^{(0)}_{\dot\alpha\dot\beta}=0$. The corresponding (chiral) strength tensor is given by
\begin{align}\label{F-inst}
(F^{(0)}_{\alpha\beta})_{ij} {}&= \epsilon^{\dot\beta\dot\gamma}\left[\partial_{(\alpha\dot\beta} A^{(0)}_{\beta)\dot\gamma}+ A^{(0)}_{(\alpha\dot\beta}  A^{(0)}_{\beta)\dot\gamma} \right]_{ij}
  =-\frac12 f(x) \big( \epsilon_{i\alpha}\epsilon_{j\beta}+ \epsilon_{i\beta}\epsilon_{j\alpha} \big)\,,
\end{align}
where $\partial_{\alpha\dot\beta} = (\sigma^\mu)_{\alpha\dot\beta}\partial_\mu$ and angular brackets denote symmetrization with respect to
indices, $A_{(\alpha\beta)} = A_{\alpha\beta} + A_{\beta\alpha}$. Here a shorthand notation was introduced for the instanton profile function
\begin{align}\label{f}
f(x) = {16\rho^2\over (x^2 + \rho^2)^2}\,.
\end{align}
It is easy to verify that $(F_{\alpha\beta})_{ij}$ is symmetric with respect to both pair of indices and satisfies the equations of motion
$D^{\dot\alpha\alpha} F_{\alpha\beta}= [\partial^{\dot\alpha\alpha}+A^{\dot\alpha\alpha}, F_{\alpha\beta} ] =0$.
Notice that \re{A-sym} and \re{F-inst} do not depend on the position of the instanton $x_0$. To restore this dependence it suffices 
to apply the shift $x\to x-x_0$.

We can further simplify \re{A-sym} and \re{F-inst} by contracting all Lorentz indices with auxiliary (commuting) spinors
$\ket{n}\equiv \lambda_\alpha$ and $|n]\equiv \bar\lambda_{\dot\alpha}$ depending on their chirality, e.g.
\begin{align}\notag
{}&  \bra{n} A_{ij} |n] \equiv\lambda^\alpha \bar\lambda^{\dot\alpha}(A_{\alpha\dot\alpha})_{ij} \,,
\qqqquad \vev{n| F_{ij} |n}  \equiv\lambda^\alpha\lambda^\beta (F_{\alpha\beta})_{ij} \,.
\end{align}
The resulting expressions only have $SU(2)$ indices and are homogenous polynomials in $\lambda$ and $\bar\lambda$.
In particular,
\begin{align}\notag\label{A-spin}
{}&  \bra{n} A^{(0)}_{ij} |n] = - { \lambda_i x_{j\dot\alpha}+\lambda_j x_{i\dot\alpha} \over x^2 + \rho^2}  \bar\lambda^{\dot\alpha}\,,
 \\
{}&   \vev{n| F^{(0)}_{ij} |n} = - f(x) \lambda_i \lambda_j \,,
\end{align}  
where the superscript `$(0)$' indicates that these expressions correspond to the lowest term in the expansion \re{dec0}.
An unusual feature of the expressions on the right-hand side of \re{A-spin} is that the chiral Lorentz indices are identified with the $SU(2)$ indices.  
 
To obtain the subleading corrections in the instanton solution \re{dec0}, depending on fermion modes, we apply the method described in Appendix~\ref{app:B}. Namely, we make use of \re{sub-fields} and replace the gauge field by its expression \re{A-spin}. Going through the calculation we
get
\begin{align}\notag\label{gauge}
{}& \bra{n} A^{(4)}_{ij} |n] =    {8\rho^2\over (\rho^2+x^2)^3} \epsilon_{ABCD} \vev{n\zeta^A} \left(\rho^2[\bar\eta^B n] - \bra{\xi^{B}}x|n]\right) \zeta_i^{C} \zeta_j^{D} \,,
\\
{}& \vev{n|F^{(4)}_{ij}|n} =  \frac14 f^2(x)   \epsilon_{ABCD} \vev{n \zeta^A}\vev{n \zeta^B}  \zeta^C_i \zeta^D_j \,,
\end{align}  
where $\zeta$ stands for a linear $x-$dependent combination of fermion modes
\begin{align}\label{xi1}
\zeta^A_\alpha(x) = \xi^A_\alpha + x_{\alpha\dot\alpha}\bar\eta^{\dot\alpha A}\,,
\end{align}
and a shorthand notation is used for various contractions of Lorentz indices
\begin{align}
\vev{n \zeta^A} = \lambda^\alpha \zeta^A_\alpha\,,\qqquad [\bar\eta^B n] = \bar\eta^B_{\dot\alpha} \bar\lambda^{\dot\alpha}\,,\qqquad 
\bra{\xi^{B}}x|n] = \xi^{\alpha B}x_{\alpha\dot\alpha} \bar\lambda^{\dot\alpha}\,.
\end{align}
Note that the dependence on the fermion modes enters into $\vev{n|F^{(4)}_{ij}|n}$ through the linear combination \re{xi1}. As explained
in Appendix~\ref{app:B}, this property can be understood using conformal symmetry. 

We recall that, to leading order, the instanton solution \re{A-spin} satisfies the self-duality condition 
$F_{\dot\alpha\dot\beta}^{(0)}=0$.  
Using the obtained expression for $A^{(4)}_{\alpha\dot\alpha}$, we find that the anti-self-dual component of the
gauge field strength tensor $[n|\bar F_{ij}|n] \equiv \bar \lambda^\alpha \lambda^\beta F_{\dot\alpha\dot\beta,ij}$  receives a  
nonvanishing correction 
\begin{align}
[n|\bar F^{(4)}_{ij}|n] = {f^2(x)\over 2\rho^2 } \epsilon_{ABCD} 
\left(\rho^2[\bar\eta^A n] - \bra{\xi^{A}}x|n]\right)\left(\rho^2[\bar\eta^B n] - \bra{\xi^{B}}x|n]\right)
 \zeta^C_i \zeta^D_j\,,
\end{align}
which contains four fermion modes. This relation illustrates that higher components of fields  
depend on fermion modes in a nontrivial manner.

Repeating the same analysis we can evaluate subleading corrections to all fields in \re{dec0}. In particular, using the relations \re{LO-fields} 
and \re{sub-fields} we get the following expressions for the scalar field on the instanton background
\begin{align}\notag\label{phi}
{}& \phi^{AB,(2)}_{ij}  = - {f(x)\over 2\sqrt 2} \zeta^{[A}_i  \zeta^{B]}_j  \,,
 \\
{}& \phi^{AB,(6)}_{ij}  = - {f^2(x)\over 20\sqrt{2}}\epsilon_{CDEF}   (\zeta^2)^{AC} (\zeta^2)^{BD}  \zeta^E_i \zeta^F_j\,,
\end{align}
where brackets in the first relation denote antisymmetrization with respect to the $SU(4)$ indices, the variable $\zeta$ is
defined in \re{xi1} and $ (\zeta^2)^{AB} = (\zeta^2)^{BA} =  \zeta^{\beta A}\epsilon_{\beta\gamma} \zeta^{\gamma B}$.
  
The expressions \re{gauge} and \re{phi} depend on the size of the instanton, $\rho$, as well as on
 $16$ fermion modes, $\xi^A_\alpha$ and $\bar\eta^{\dot\alpha A}$. To restore the dependence on the position of the instanton,
 we apply the shift $x\to x-x_0$. 
 
\section{Correlation functions in the semiclassical approximation}  \label{sect:semi}
 
In the semiclassical approximation, the calculation of correlation functions reduces to averaging the classical profile of the operators over 
the collective coordinates of instantons \cite{Belitsky:2000ws,Dorey:2002ik,Bianchi:2007ft}  
\begin{align}\label{corr}
\vev{O_1\dots O_n}_{\rm inst} = \int d\mu_{\rm phys} \e^{-S_{\rm inst}}  O_1\dots O_n\,,
\end{align}
where the gauge invariant operators $O_i$ on the right-hand side are evaluated on the instanton background \re{dec0}. For the $SU(2)$ gauge group,
the collective coordinates of the one-instanton solution are the size of the instanton $\rho$, its localtion $x_0^\mu$ and $16$ fermion
modes, $\xi_\alpha^A$ and $\bar\eta_{\dot\alpha}^A$. The corresponding integration measure
is \cite{Bianchi:1998nk}
\begin{align}\label{measure}
 \int d\mu_{\rm phys} \e^{-S_{\rm inst}} = {g^8 \over 2^{34}\pi^{10}}\e^{2\pi i\tau} \int d^4 x_0 \int {d\rho\over\rho^5} \int d^8 \xi
 \int d^8\bar\eta\,,
\end{align}
where the complex coupling constant  $\tau$ is defined in \re{tau}. 

We recall that, due to our normalization of the Lagrangian \re{La}, the instanton background \re{dec0} does not depend on the coupling constant. The same 
is true for the operators \re{1/2}, \re{O-LC} and \re{O-sum0} built from scalar and gauge fields. As a consequence, the instanton correction to the correlation
function of these operators scales in the semiclassical approximation as $O(g^8\e^{2\pi i\tau})$ independently on the number of operators $n$. At the same time,
the same correlation functions in the Born approximation scale as $O(g^{2n})$ with one power of $g$ coming from each scalar field (see Eqs.~\re{3pt-Born}
and \re{2pt-Born}).~\footnote{In the previous paper \cite{Alday:2016tll}, we used different normalization of operators, $O_{\rm there} = O_{\rm here}/g^2$, 
for which the correlation functions in the Born approximation do not depend on the coupling constant.
}
Thus, the ratio of the two contributions scales as $O(g^{8-2n}\e^{2\pi i\tau})$ with $n$ being the number of operators.  As we show in a moment, the instanton correction vanishes in the semiclassical approximation
for $n>5$. Later in the paper
we shall compute this correction explicitly for $n=2$ and $n=3$. 
 
For the correlation function \re{corr} to be different from zero upon integration of fermion modes, the product of operators $O_1\dots O_n$ should soak up  the product of all fermion modes $(\xi)^8 (\bar\eta)^8 \equiv \prod_A \xi^A_1\xi^A_2  \bar\eta_{\dot 1}^A\bar\eta_{\dot 2}^A$. Let us denote by $k_i$ the minimal number of fermion modes in $O_i$. 
Then, by virtue of the $SU(4)$ invariance, the total number of modes in the product $O_1\dots O_n$ is given by 
$k_{\rm min}+4p$ with $k_{\rm min} = k_1+\dots +k_n$ and $p=0,1,\dots$. For $k_{\rm min} >16$ the product $O_1\dots O_n$ is necessarily proportional to the square of  a fermion mode and, therefore, the correlation function \re{corr} vanishes. To obtain a nontrivial result for the correlation function, we have to go beyond the semiclassical approximation
and take into account quantum fluctuations of fields. For $k_{\rm tot}\le 16$ and $k_{\rm min}$ multiple of $4$,  the integral over fermion modes in \re{corr}
does not vanish a priori. This case corresponds to the so-called minimal correlation functions \cite{Green:2002vf,Bianchi:2007ft}. Another interesting feature of these correlation functions 
is that  the results obtained 
in the semiclassical approximation for the $SU(2)$ gauge group can be extended to the general case of $SU(N)$ gauge group 
for the one-instanton 
solution \cite{Dorey:1998xe,Dorey:2002ik} and for multi-instanton solutions at large $N$ \cite{Dorey:1999pd}.

Since the operators \re{1/2}, \re{O-LC} and \re{O-sum0} involve two scalar fields, we deduce from \re{dec0} that each of them has at least four fermion modes,
$k=4$. Then, for the product of these operators $O_1\dots O_n$ we have $k_{\rm min} = 4n$ and, as a consequence, the correlation function
$\vev{O_1\dots O_n}_{\rm inst}$ is minimal for $n\le 4$.

\subsection{Instanton profile of twist-two operators} 
 
To compute the correlation functions \re{gen-exp} in the semiclassical approximation, we have to evaluate the operators $O_{\bf 20'}$ and $O_S$ on the instanton 
background by replacing all fields by their explicit expressions \re{dec0}
and, then, expand their product in powers of $16$ fermion modes. 

For the half-BPS operators \re{1/2} we find \cite{Alday:2016tll}
\begin{align}\label{1/2BPS}
O^{(4)}_{\bf 20'}(x) = \frac12 f^2(x) Y_{AB} Y_{CD} (\zeta^2)^{AC} (\zeta^2)^{BD}\,,
\end{align} 
where $f(x)$ is the instanton profile function \re{f} and the superscript on the left-hand side indicates the number of fermion modes.
Since $O_{\bf 20'}(x)$ is annihilated by half of the $\mathcal N=4$ supercharges, its expansion involves only four fermion modes. 

As argued above, in order to study correlators involving  twist-two operators it is convenient to introduce light-ray operators \re{O-LC}. In order to evaluate those in the instanton background we replace scalar and gauge fields by their expressions $Z= Z^{(2)} + Z^{(6)} + \dots$ and
$A=A^{(0)} + A^{(4)} + \dots$, which leads to
\begin{align}\label{OO-dec}
\mathbb O(z_1,z_2) = \mathbb O^{(4)} + \mathbb O^{(8)} +\mathbb O^{(12)} +\mathbb O^{(16)} \,,
\end{align} 
where the leading term contains four fermion modes and each subsequent term has four modes more. The last term  
is proportional to the product of all $16$ modes $(\xi)^8 (\bar\eta)^8$. 
The first two terms on the right-hand side of \re{OO-dec} are given by 
\begin{align}\notag\label{O4-8}
\mathbb O^{(4)}(z_1,z_2) {}&=  \tr\Big[Z^{(2)} (nz_1) E^{(0)} (z_1,z_2) Z^{(2)} (nz_2) E^{(0)}(z_2,z_1)\Big]\,,
\\ \notag
\mathbb O^{(8)}(z_1,z_2) {}&=  \tr\Big[Z^{(6)} (nz_1) E^{(0)} (z_1,z_2) Z^{(2)} (nz_2) E^{(0)}(z_2,z_1)\Big]
\\
{}& + \tr\Big[Z^{(2)} (nz_1) E^{(4)} (z_1,z_2) Z^{(2)} (nz_2) E^{(0)}(z_2,z_1)\Big]
+ (z_1 \leftrightarrow z_2)\,,
\end{align}
where $E^{(n)}$ denote specific terms in the expansion of the light-like Wilson line \re{line} on the instanton background
\begin{align}
E(z_1,z_2) = E^{(0)} + E^{(4)} + \dots\,.
\end{align}
Explicit expressions for $E^{(0)}$ and $E^{(4)}$ can be found in Appendix~\ref{App:WL}. 

The remaining terms $\mathbb O^{(12)}$ and $\mathbb O^{(16)}$ are given by a priori lengthy expressions but
their evaluation leads to a surprisingly simple result
\begin{align}\label{O-zero}
\mathbb O^{(12)}(z_1,z_2) = \mathbb O^{(16)}(z_1,z_2) = 0\,.
\end{align}
As we show in Appendix~\ref{App:WL}, these terms are necessarily proportional to the square of a fermion mode and, therefore, have to vanish.

Let us consider the correlation function \re{G4-lc}. As was explained at the beginning of the section, this correlation function is minimal and scales as
$O(g^8\e^{2\pi i \tau})$ in the semiclassical approximation. Applying \re{corr}, we have to find the instanton profile of 
$\mathbb O(z_1,z_2)  O_{\bf 20'}(x_1)O_{\bf 20'}(x_2)$ and identify the contribution containing $16$ fermion modes. Making use of \re{1/2BPS} and \re{OO-dec} we
find
\begin{align}\label{start1}
\vev{\mathbb O(z_1,z_2)  O_{\bf 20'}(x_1)O_{\bf 20'}(x_2)}_{\rm inst} = \int d\mu_{\rm phys} \e^{-S_{\rm inst}}  \mathbb O^{(8)}(z_1,z_2)O_{\bf 20'}^{(4)}(x_1)O_{\bf 20'}^{(4)}(x_2)\,.
\end{align}
Later we shall use this relation to find the instanton
correction to the three-point correlation function $\vev{ O_S(0) O_{\bf 20'}(x_1)O_{\bf 20'}(x_2)}_{\rm inst}$. 

In a similar manner, in order to find $\vev{ O_S(0)\bar O_{S'}(x)}_{\rm inst}$ we have to consider the correlation function of two light-ray operators separated 
by a distance $x$ and conjugated to each other. For this purpose, we have to generalize the definition \re{O-LC} by allowing the light-ray to pass through an 
arbitrary point $x$
\begin{align}\label{shift}
\mathbb O(z_1,z_2|x) = \e^{i P \cdot x} \mathbb O(z_1,z_2)  \e^{-i P\cdot x}\,,
\end{align}
where $P_\mu$ is the operator of total momentum. The operator \re{shift} is given by the same expression \re{O-LC} with scalar and gauge fields shifted by $x$.
Obviously, for $x=0$ we have $\mathbb O(z_1,z_2|0)= \mathbb O(z_1,z_2)$. Substituting \re{OO-dec} into \re{shift}, we find that $\mathbb O(z_1,z_2|x)$ 
has a similar expansion in powers of fermion modes. Then, we take into account \re{O-zero} to get
\begin{align}\label{start2}
\vev{\mathbb O(z_1,z_2|0)  \bar{\mathbb O}(z_3,z_4|x) }_{\rm inst} = \int d\mu_{\rm phys} \e^{-S_{\rm inst}}  \mathbb O^{(8)}(z_1,z_2|0)
\bar{\mathbb O}^{(8)}(z_3,z_4|x) \,,
\end{align}
where $\bar{\mathbb O}(z_3,z_4|x)$ is a conjugated operator.
Notice that the lowest $\mathbb O^{(4)}$ term of the expansion \re{OO-dec} does not contribute to \re{start1} and \re{start2}.

We are now ready to compute the correlation functions \re{start1} and \re{start2}. Using the expression for the integration measure \re{measure}, we first
perform the integration over fermion modes. Replacing $O^{(4)}_{\bf 20'}$ and $\mathbb O^{(8)}$ by their explicit expressions, Eqs.~\re{1/2BPS} and \re{O4-8}, respectively, and going through a lengthy calculation we find (see Appendix~\ref{app:der1} for details)
\begin{align}\notag\label{aux3}
\int d^8\xi d^8\bar\eta\,  \mathbb O^{(8)}(z_1,z_2)O_{\bf 20'}^{(4)}(x_1)O_{\bf 20'}^{(4)}(x_2) =   -   2^{29} \times 3^3 \times (Y_1 Y_2) (Y_1 Y_Z)(Y_2 Y_Z)
\\
 \times{(z_1-z_2)^2 \,  \big[(nx_2)x_{1}^2-(nx_1)x_{2}^2\big]^2 \rho^{14} \over [(\rho^2 + (nz_1 -x_0)^2)  (\rho^2 +  (nz_2 -x_0)^2)]^3 [(\rho^2+x_{10}^2)  (\rho^2+x_{20}^2)]^4}  \,,
\end{align}
where $(Y_i Y_j) = \epsilon^{ABCD} Y_{i,AB} Y_{i,CD}$ and the antisymmetric tensors $Y_{i,AB}$ carry the $SU(4)$ charges of the operators. 
The tensors $Y_{i,AB}$ (with $i=1,2$) enter into the definition \re{1/2} of the operators $O_{\bf 20'}(x_i)$ and $Y_{Z,AB}$ defines the complex scalar field
$Z=\phi^{14}= Y_{Z,AB} \phi^{AB}$. In a similar manner, we find from \re{start2}  (see Appendix~\ref{app:der2} for details)
\begin{align}\notag\label{aux2}
{}& \int d^8\xi d^8\bar\eta\, \mathbb O^{(8)}(z_1,z_2|0)
\bar{\mathbb O}^{(8)}(z_3,z_4|x)  
\\
{}& =   { 2^{30}\times 3^2\times (z_1-z_2)^2 (z_3-z_4)^2  (xn)^4 \rho^{12} \over [(\rho^2 + (nz_1 -x_0)^2) (\rho^2 +  (nz_2 -x_0)^2)(\rho^2 + (x+nz_3 -x_0)^2)(\rho^2 +  (x+nz_4 -x_0)^2)]^3}
\end{align}
Notice that \re{aux3} and \re{aux2} vanish quadratically for $z_1\to z_2$ (as well as for $z_3\to z_4$). This property can be understood as follows. For $z_2\to z_1$ we can use the definition of the light-ray operator \re{O-LC} to show that
\begin{align}\label{O-zero1}
\mathbb O(z_1,z_2) =  \tr\big[Z^2(nz_1)\big] + \frac12 (z_2-z_1) \partial_+ \tr\big[Z^2(nz_1)\big]  + O\left((z_2-z_1)^2\right)\,.
\end{align}
The first two terms of the expansion involve the half-BPS operator $\tr Z^2$, it can be obtained from the general expression \re{1/2} for $Y=Y_Z$.
According to \re{1/2BPS}, the half-BPS operators have exactly four fermion modes and, therefore, the first two
terms on the right-hand side of \re{O-zero1} do not contribute to $\mathbb O^{(8)}(z_1,z_2)$ leading to $\mathbb O^{(8)}(z_1,z_2)=O((z_2-z_1)^2)$.

To obtain the correlation functions, we have to integrate \re{aux3} and \re{aux2} over the bosonic coordinates $\rho$ and $x_0$
with the measure \re{measure}. The resulting integrals
can be expressed in terms of the so-called $D-$functions
\begin{align}\label{D-fun}
D_{\Delta_1\dots\Delta_n}(x_1,\dots,x_n) = \int d^4 x_0 \int {d\rho\over\rho^5} \prod_{i=1}^n  \left( {\rho \over \rho^2 + x_{i0}^2} \right)^{\Delta_i},
\end{align}
where $x_{i0} = x_i-x_0$. Multiplying  \re{aux3} and \re{aux2} by the additional factors
 coming from the integration measure \re{measure}, we finally find the leading 
instanton correction to the correlation functions
\begin{align}\notag\label{3pt-inst}
{}&  \vev{\mathbb O(z_1,z_2) O_{\bf 20'}(1) O_{\bf 20'}(2)}_{\rm inst}   = -{27\over 128 \pi^{10}} 
 g^{8}  \e^{2\pi i \tau} (Y_1 Y_2) (Y_1 Y_Z)(Y_2 Y_Z)
 \\ 
{}& \hspace*{20mm}\times
 (z_1-z_2)^2 (x_1^2 x_2^2)^2 \left[{2(nx_1)\over x_{1}^2}-{2(nx_2)\over x_{2}^2}\right]^2
 D_{3344}(nz_1,nz_2,x_1,x_2)\,,
\\[2mm] \notag
 {}& \vev{\mathbb O(z_1,z_2|0)\bar{\mathbb O}(z_3,z_4|x) } _{\rm inst} ={9\over 16 \pi^{10}}
g^{8}  \e^{2\pi i \tau}
\\ \label{2pt-inst}
{}& \hspace*{20mm}\times
(z_1-z_2)^2 (z_3-z_4)^2  (xn)^4
 D_{3333}(nz_1,nz_2,x+nz_3,x+nz_4)\,.
\end{align}
The same correlation functions in the Born approximation are given by \re{OOO-B} and \re{OO-B}, respectively, evaluated for $N=2$. Dividing \re{3pt-inst}
and \re{2pt-inst} by the Born level expressions, we find that the instanton corrections to the two correlation functions scale as 
$O(g^2 \e^{2\pi i \tau})$ and $O(g^4 \e^{2\pi i \tau})$, respectively, in agreement with the analysis at the beginning of this section.

\subsection{Instanton corrections to correlation functions} 

Let us apply \re{3pt-inst} and \re{2pt-inst} to derive the correlation functions of twist-two operators. We recall that to the 
lowest order in the coupling constant, the correlation functions involving light-ray and twist-two operators are related to each other 
through the relations \re{double} and \re{use1}. It is not obvious however whether the same relations should work for the instanton 
corrections in the semiclassical approximation. To show that this is the case, we apply below \re{double} and \re{use1} to relations \re{3pt-inst} and \re{2pt-inst} and 
verify that the obtained expressions for the correlation functions $\vev{O_S(0)O_{\bf 20'}(1) O_{\bf 20'}(2)}$ and $\vev{O_S(0) O_{S'}(x)}$
are indeed consistent with conformal symmetry and have the expected form \re{gen-exp}.

Applying \re{use1} to \re{3pt-inst}, we get
\begin{align}\notag\label{O_S-I_S}
\vev{O_S(0) O_{\bf 20'}(1) O_{\bf 20'}(2)}_{\rm inst} {}& = -{27\over 128 \pi^{10}} 
 g^{8}  \e^{2\pi i \tau} (Y_1 Y_2) (Y_1 Y_Z)(Y_2 Y_Z)
 \\ 
{}& \times
  (x_1^2 x_2^2)^2 \left[{2(nx_1)\over x_{1}^2}-{2(nx_2)\over x_{2}^2}\right]^2
I_S(x_1,x_2)\,,
\end{align}
where $I_S(x_1,x_2)$ denotes the following integral
\begin{align}\label{aux-int}
 I_S(x_1,x_2)= \oiint {dz_1dz_2\over (2\pi i)^2} {(z_1-z_2)^S\over (z_1z_2)^{S+1}}   (z_1-z_2)^2 D_{33
 \Delta\Delta}(nz_1,nz_2,x_1,x_2)\,,
\end{align}
with $\Delta=4$. In what follows we relax this condition and treat $\Delta$ as an arbitrary parameter. The reason for this is that the same integral with $\Delta=3$ enters the calculation 
of \re{2pt-inst}. 

Replacing the $D-$function in \re{aux-int} by its integral representation \re{D-fun} and exchanging the order of integration, we find that the integrand has triple poles at $z_1,z_2=(\rho^2 +x_0^2)/(2(x_0n))$. Blowing up the integration contour in \re{aux-int} and
picking up the residues at these poles, we find that the integral over $z_1$ and $z_2$ vanishes for all nonnegative integer $S$ except for
$S=2$, leading to
\begin{align}\notag\label{I_S}
I_S(x_1,x_2)
{}&= 6 \, \delta_{S,2} D_{6\Delta\Delta}(0,x_1,x_2)
\\
{}&= 6 \, \delta_{S,2}\int   
{d^4 x_0 \, d\rho\,\rho^{9}\over (\rho^2+x_0^2)^6 (\rho^2+x_{10}^2)^\Delta (\rho^2+x_{20}^2)^\Delta} \,.
\end{align}
For $\Delta=4$ the calculation of this integral yields 
\begin{align}\label{D644}
D_{644}(0,x_1,x_2) = {\pi^2\over 90} {1\over x_{12}^2 (x_1^2x_2^2)^3}\,.
\end{align}
Substituting \re{I_S} and \re{D644} into \re{O_S-I_S} we finally obtain
\begin{align}\notag \label{main1}
\vev{O_S(0) O_{\bf 20'}(1) O_{\bf 20'}(2)}_{\rm inst} {}& = - \delta_{S,2} {9\over 640 \pi^8} 
 g^{8}  \e^{2\pi i \tau} (Y_1 Y_2) (Y_1 Y_Z)(Y_2 Y_Z)
 \\ 
{}& \times
  {1\over x_{12}^2 x_1^2 x_2^2} \left[{2(nx_1)\over x_{1}^2}-{2(nx_2)\over x_{2}^2}\right]^2\,.
\end{align}
Surprisingly enough, this expression vanishes for all spins except $S=2$. For $S=0$ the corresponding
twist-two operator $O_{S=0} = \tr Z^2$ is half-BPS. In this case, the three-point correlation function is protected 
from quantum corrections and, therefore, the instanton correction \re{main1} should vanish for $S=0$.

To obtain the two-point correlation function $\vev{O_S(0) O_{S'}(x)}$ from \re{2pt-inst} we apply \re{double} 
to both operators to get
\begin{align}
\vev{O_S(0) O_{S'}(x)}_{\rm inst} = {9\over 16 \pi^{10}}  
g^{8}  \e^{2\pi i \tau}  (xn)^4 I_{SS'}(x)\,,
\end{align}
where $I_{SS'}(x)$ is given by a folded contour integral over four $z-$variables. Using \re{aux-int} this integral can be rewritten as
\begin{align}
I_{SS'}(x)= \oiint {dz_3dz_4\over (2\pi i)^2} {(z_3-z_4)^{S'+2}\over (z_3z_4)^{S'+1}}   
I_S(x+nz_3,x+nz_4)\,,
\end{align}
where $I_S$ is evaluated for $\Delta=3$. Replacing $I_S$ with its integral representation \re{I_S}, we evaluate
the integral by picking up the residue at the poles $z_3,z_4=(\rho^2 +(x_0-x)^2)/(2(x_0-x)n)$ in the same manner as
\re{I_S}. In this way, we arrive at
\begin{align}\label{OO-div}
\vev{O_S(0) \bar O_{S'}(x)}_{\rm inst} =  \delta_{S,2} \delta_{S,S'} {81\over 4 \pi^{10}} g^{8}  \e^{2\pi i \tau}  (xn)^4 D_{66}(0,x)\,,
\end{align}
where the $D-$function is defined in \re{D-fun}. Similar to \re{main1}, this expression vanishes for all spins except
$S=S'=2$. For $S=S'=0$ this property is in agreement with protectiveness of two-point correlation
functions of half-BPS operators.

A close examination of \re{D-fun} shows that $D_{66}(0,x)$ develops a logarithmic
divergence. It comes from integration over small size instantons, $\rho\to 0$, located close to one of the operators,
$x_0^2\to 0$ and $(x-x_0)^2\to 0$. This divergence produces a logarithmically enhanced contribution $\sim \ln x^2$
which modifies the scaling dimensions of twist-two operators. To identify this contribution, we regularize the integral
by modifying the integration measure over $x_0$ 
\footnote{We would like to emphasize that this regularization is different from the conventional
dimensional regularization. Since the coefficient in front of $\ln x^2$ in \re{D-reg} 
is independent on the choice of regularization, we can choose it to our convenience. }
\begin{align}\notag\label{D-reg}
D_{66}(0,x) {}& \to \int d^{4-2\epsilon} x_0 \int {d\rho\over\rho^5} {\rho^{12}\over (\rho^2+x_0^2)^6 (\rho^2+(x-x_0)^2)^6}
\\ {}&
= {\pi^{2}\over 20(x^2)^6} \left(-{1\over \epsilon} + \ln x^2  + \dots \right)\,,
\end{align}
where dots denote terms independent of $x$ and/or vanishing for $\epsilon\to 0$. In the standard manner, the pole $1/\epsilon$ can
be removed from the correlation function $\vev{O_S(0) O_{S'}(x)}$ by multiplying the twist-two operators, $O_S(x) \to Z(1/\epsilon) O_S(x)$ by appropriately chosen renormalization $Z-$factor.
Retaining only this term we get from
\re{OO-div}
\begin{align}\label{main2}
\vev{O_S(0) \bar O_{S'}(x)}_{\rm inst} =  \delta_{S,2} \delta_{S,S'}{81\over 80 \pi^8} g^{8} \e^{2\pi i \tau}  { (xn)^4\over (x^2)^6}  \ln x^2\,.
\end{align}
Relations \re{main1} and \re{main2} define one-instanton corrections to the correlation functions. Anti-instanton corrections are given by the complex conjugated expressions.

Finally, we combine \re{main1} and \re{main2} with analogous expressions in the Born approximation (given by \re{3pt-Born} and 
\re{2pt-Born} for $N=2$) and add the anti-instanton contribution to get the following expressions for the correlation functions
in $\mathcal N=4$ SYM for the 
$SU(2)$ gauge group
\begin{align}\notag\label{3pt-I}
{}& \vev{O_S(0) O_{\bf 20'}(1) O_{\bf 20'}(2)}  = {3 g^6\over (4\pi^2)^3} {(Y_1Y_2) (Y_1 Y_Z) (Y_2 Y_Z) \over x_{12}^2 x_1^2 x_2^2}
\left[{2(nx_1)\over x_1^2} -{2(nx_2)\over x_2^2}\right]^S 
\\
{}& \hspace*{40mm} \times \left[ 1- \delta_{S,2} {3g^2 \over 10\pi^2} ( \e^{2\pi i \tau}+ \e^{-2\pi i \tau})\right],
\\[2mm] \notag\label{2pt-I}
{}& \vev{O_S(0) \bar O_{S'}(x)} =  \delta_{SS'} {3 g^4 c_{S} \over 2 (4\pi^2)^2}{\left[2(nx)\right]^{2S}\over (x^2)^{2+2S}} 
\\
{}& \hspace*{40mm} \times 
 \left[ 1+\delta_{S,2} {9\over 5} \lr{g^2 \over 4\pi^2}^2 ( \e^{2\pi i \tau}+ \e^{-2\pi i \tau}) \ln x^2 \right],
\end{align}
where the normalization factor $c_S$ is defined in \re{2pt-Born}. We recall that these relations were derived in the one (anti)instanton sector
in the semiclassical
approximation and are valid up to corrections suppressed by powers of $g^2$.

Comparing \re{2pt-I} with the general expression for a two-point correlation function \re{gen-exp}, we obtain the leading 
instanton correction to the scaling dimension of twist-two operators $\Delta_S = 2 + S +\gamma_S$
\begin{align}\label{anom} 
 \gamma_{S}^{\rm (inst)} =  -\delta_{S,2} {9\over 5} \lr{g^2 \over 4\pi^2}^2 ( \e^{2\pi i \tau}+ \e^{-2\pi i \tau})\,.
\end{align}
In a similar manner, matching \re{3pt-I} and \re{gen-exp} we obtain the following result for the properly normalized OPE coefficient 
$\widehat C_S= C_S/C_{S}^{\rm (Born)}$
\begin{align}\label{C-hat}
\widehat C_S =  1- \delta_{S,2} {3g^2 \over 10\pi^2} ( \e^{2\pi i \tau}+ \e^{-2\pi i \tau})\,.
\end{align}
As follows from \re{2pt-I}, the instanton correction to the normalization factor $\mathcal N_S$ entering the first relation in \re{gen-exp} 
has the same dependence on the coupling constant as \re{anom} and does not affect the leading correction to the 
structure constants in the OPE of two half-BPS operators.

As was already mentioned, for $S=0$ the correlation functions \re{3pt-I} and \re{2pt-I} are protected from quantum corrections
and, therefore, $\gamma_{S=0}=0$ and $\widehat C_{S=0} = 1$ for arbitrary coupling constant.
For $S>2$, there are
no reasons for the same relations to hold beyond the semiclassical approximation.
For $S=2$, the corresponding conformal operator
$O_{S=2} \sim \tr(ZD_+^2 Z) - 2 \tr(D_+Z D_+Z)$  belongs to the same $\mathcal N=4$ supermultiplet
as Konishi operator  $K= \tr(\phi^{AB} \bar \phi_{AB})$. As a consequence, the two operators should
have the same anomalous dimension, $\gamma_{S=2}=\gamma_K$, as well as OPE coefficients,
$\widehat C_{S=2}=\widehat C_K$. As a nontrivial check of our calculation we use the results 
of \cite{Alday:2016tll} to verify that both relations are indeed satisfied in the semiclassical approximation.

As was already mentioned in Section \ref{sect:semi}, the correlation functions \re{3pt-I} and \re{2pt-I} 
belong to the class of minimal correlation functions. Following \cite{Dorey:1998xe,Dorey:1999pd,Dorey:2002ik}, we can then generalize the
relations \re{3pt-I} and \re{2pt-I} to the $SU(N)$ gauge group and, in addition, include the contribution 
of an arbitrary number of (anti)instantons at large $N$. As before, for $S\neq 2$ the resulting expressions 
for $\gamma_S$ and $\widehat C_S$ do not receive corrections in the semiclassical approximation, 
whereas for $S=2$ they coincide with those for the Konishi operator and can be found in \cite{Alday:2016tll}.

\section{Conclusions} 

In this work we have presented the explicit calculation of the instanton contribution to two-
and three-point correlation functions involving half-BPS and twist-two operators. 
A somewhat surprising outcome of our analysis is the vanishing of the leading instanton corrections to the scaling
dimensions and the OPE coefficients of twist-two operators with spin $S>2$. This result comes from a rather involved 
calculation and requires a better understanding. Note that the situation here is very different from that for twist-four operators. As it was shown in \cite{Alday:2016tll}, crossing symmetry implies that twist-four operators with arbitrarily high spin acquire instanton corrections already 
at order $O(g^2\e^{-8\pi^2/g^2})$. 

According to \re{res0}, the instanton corrections to twist-two operators with $S>2$ are pushed to higher order in $g^2$. Their calculation remains
a challenge as it requires going beyond the semiclassical approximation. There is however an interesting high spin limit, $S\gg 1$,
in which we can get additional insight on the instanton effects. In this limit, the scaling dimensions of
twist-two operators scale logarithmically with the spin \cite{Korchemsky:1988si,Alday:2007mf}
\begin{align}
\Delta_S = S+2\Gamma_{\rm cusp}(g^2) \ln S +O(S^0)\,,
\end{align}
where $\Gamma_{\rm cusp}(g^2)$ is the cusp anomalous dimension. The instanton contribution to $\Delta_S$ should
have the same asymptotic behavior and produce a correction to $\Gamma_{\rm cusp}(g^2)$. According to the first
relation in \re{res0}, it should scale at least as $O(g^6\e^{-8\pi^2/g^2})$. We can find however the same correction 
using the fact that the cusp anomalous dimension governs
the leading UV divergences of light-like polygon Wilson loops.  

The light-like polygon Wilson loop $W_L$ is given by the product of gauge links \re{line} defined for $L$ different light-like vectors $n_i$.
In the semiclassical approximation, the instanton contribution to $W_L$ can be found using \re{corr}. Since $W_L$ does not depend on the
coupling constant on the instanton background, the dependence on $g^2$ only comes from the integration measure \re{measure}
leading to $\vev{W_L}_{\rm inst} =O(g^8\e^{-8\pi^2/g^2})$. At the same time,  in the Born approximation, we have
$\vev{W_L}_{\rm Born} =1$. As a consequence, the leading instanton correction to the cusp anomalous dimension 
scales at least as  
\begin{align}\label{cusp}
\Gamma_{\rm cusp}^{\rm (inst)}(g^2)=O(g^8\e^{-8\pi^2/g^2})\,.
\end{align}
Combining the last two relations we conclude that  $\Delta_S^{\rm (inst)}= O(g^8\e^{-8\pi^2/g^2}\ln S)$ at large $S$. Thus, if
the leading $ O(g^6 \e^{-8\pi^2/g^2})$ correction to $\Delta_S^{\rm (inst)}$ in \re{res0} is different from zero, it should approach a finite value for 
$S\to \infty$.~\footnote{Similar considerations should apply to the structure constants $C_S^{\rm (inst)}$ in the large spin limit. Indeed, from the analysis of \cite{Alday:2013cwa} it follows that the structure constants for $S\gg 1$ are given in terms of the cusp anomalous dimension.}
It would be interesting to compute explicitly the instanton correction \re{cusp}. 

\section*{Acknowledgements}  

We are grateful to Massimo Bianchi, Emeri Sokatchev, Yassen Stanev and Pierre Vanhove for useful discussions.   The work of L.F.A. was supported by ERC STG grant 306260. L.F.A. is a Wolfson Royal Society Research Merit Award holder.  This work of G.P.K. was supported in part by
the French National Agency for Research (ANR) under contract StrongInt (BLANC-SIMI-
4-2011).
 
\appendix
   
\section{$\mathcal N=4$ SYM in spinor notations}\label{app:A}

Performing the calculation of instanton corrections in $\mathcal N=4$ SYM it is convenient to employ spinor notations. We use Pauli matrices $\sigma^\mu = (1, i\boldsymbol \sigma)$ to map an arbitrary four-dimensional Euclidean vector $x_\mu$ into a
$2\times 2$ matrix
\begin{align}
x_{\alpha\dot\beta} = x_\mu (\sigma^\mu)_{\alpha\dot\beta}\,,
\end{align} 
and use the completely antisymmetric tensor to raise and lower its indices
\begin{align}\label{raise-lower}
x^\alpha_{\dot\beta} =\epsilon^{\alpha\beta}x_{\beta\dot\beta} \,,\qqquad
x_{\alpha}^{\dot\beta} =x_{\alpha\dot\alpha}\epsilon^{\dot\alpha\dot\beta} \,,\qqquad
x^{\dot\alpha\beta} = \epsilon^{\beta\alpha}x_{\alpha\dot\beta} 
\epsilon^{\dot\beta\dot\alpha}\,,
\end{align}
with $\epsilon_{\alpha\beta} \epsilon^{\alpha\gamma} = \delta_\beta^\gamma$,
$\epsilon_{\dot\alpha\dot\beta} \epsilon^{\dot\alpha\dot\gamma} = \delta_{\dot\beta}^{\dot\gamma}$ and 
$\epsilon_{12}=\epsilon^{12}=1$. Then,
\begin{align}
x_\mu^2 = \frac12 x_{\dot\alpha}^\alpha x^{\dot\alpha}_\alpha = - \frac12 x_{\alpha\dot\alpha}x^{\dot\alpha\alpha}
=\frac12 x_{\alpha\dot\alpha}x_{\beta\dot\beta}\epsilon^{\alpha\beta}\epsilon^{\dot\alpha\dot\beta}\,.
\end{align}
For derivatives we have similarly
\begin{align}
 \partial_{\alpha\dot\beta} =  \partial_\mu(\sigma^\mu)_{\alpha\dot\beta}\,,\qqquad
 \partial_{ \alpha\dot\alpha} x_{\beta\dot\beta} =2\epsilon_{\alpha\beta}\epsilon_{\dot\alpha\dot\beta}\,,\qqquad
 \partial_{\alpha\dot\alpha} x^2 = 2 x_{ \alpha\dot \alpha}\,.
\end{align}
Throughout the paper we use the following conventions for contracting
Lorentz indices in the product of $2\times 2$ matrices  
\begin{align}
(x y)_{\alpha\beta} = x_{\alpha\dot\alpha} y^{\dot\alpha}_\beta\,,\qqquad
(x y)_{\dot\alpha\dot\beta} = x^{\alpha}_{\dot\alpha} y_{\alpha\dot\beta}\,,\qqquad
 (x y z)_{\alpha\dot\beta} = x_{\alpha\dot\alpha} y^{\dot\alpha \beta} z_{\beta\dot\beta}  \,.
\end{align}
Using these definitions we obtain for the gauge field $A_\mu$ and the stress tensor $F_{\mu\nu} = -i [D_\mu,D_\nu]$  
\begin{align}
A_{\alpha\dot\alpha} = i A_\mu (\sigma^\mu)_{\alpha\dot\alpha}\,,\qqquad
F_{\alpha\beta} =- i F_{\mu\nu} (\sigma^{\mu }\sigma^{ \nu})_{\alpha\beta} \,,\qqquad
F_{\dot\alpha\dot\beta} =- i F_{\mu\nu} (\sigma^{\mu }\sigma^{ \nu})_{\dot\alpha\dot\beta}\,,
\end{align}
where the additional factor of  $i$ is introduced for convenience.
The symmetric matrices $F_{\alpha\beta}$ and $F_{\dot\alpha\dot\beta}$ describe (anti)self-dual parts
of the strength tensor   
\begin{align}\notag
{}& F_{\alpha\beta} = \epsilon^{\dot\alpha\dot\beta} \lr{ D_{\alpha\dot\alpha} D_{\beta\dot\beta} + D_{\beta\dot\alpha} D_{\alpha\dot\beta} }\equiv  -D^2_{(\alpha\beta)}\,,
\\
{}& F_{\dot\alpha\dot\beta} =   \epsilon^{\alpha\beta} \lr{ D_{\alpha\dot\alpha} D_{\beta\dot\beta} + D_{\beta\dot\alpha} D_{\alpha\dot\beta} } \equiv  -D^2_{(\dot\alpha\dot\beta)}\,,
\end{align}
where angular brackets denote symmetrization with respect to indices and the covariant derivative $D_{\alpha\dot\alpha } =D_\mu (\sigma^\mu)_{\alpha\dot\alpha }$ is defined as
\begin{align}\label{cov}
D_{\alpha\dot\beta} X = [\partial_{\alpha\dot\beta} + A_{\alpha\dot\beta} , X]\,.
\end{align} 
The Lorentz indices are raised and lowered according to \re{raise-lower}.

The Lagrangian of $\mathcal N=4$ super Yang-Mills theory takes the following form in spinor notations 
 \begin{align}\notag\label{La}
L {}& = {1\over g^2} \tr \Big\{ - \frac1{16} F_{\alpha\beta}^2 - \frac1{16} F_{\dot\alpha\dot\beta}^2 - \frac14 D^\alpha_{\dot\alpha} \phi^{AB} D_\alpha^{\dot\alpha} \bar\phi_{AB} 
- 2 i \bar\lambda_{\dot\alpha A} D^{\dot\alpha\beta} \lambda_\beta^A+ \sqrt{2} \lambda^{\alpha A} [\bar\phi_{AB},\lambda_\alpha^B]
\\[2mm]  
{}& 
- \sqrt{2} \bar\lambda_{\dot\alpha A} [\phi^{AB},\bar\lambda_B^{\dot\alpha}]+\frac18 [\phi^{AB},\phi^{CD}][\bar\phi_{AB},\bar\phi_{CD}] \Big\}
 +  i{\theta\over 8\pi^2}\tr \Big\{ \frac1{16} F_{\alpha\beta}^2- \frac1{16} F_{\dot\alpha\dot\beta}^2\Big\}\,,
\end{align} 
where gaugino fields $\lambda_\alpha^A$ and $\bar\lambda^{\dot\alpha}_A$, and scalar fields, $\phi^{AB}$ and $\bar\phi_{AB}$,  
carry the $SU(4)$ indices ($A,B=1,\dots,4$) and  satisfy the reality condition $\bar\phi_{AB} = \frac12\epsilon_{ABCD} \phi^{CD}$. All fields are in the adjoint representation of  the $SU(N)$ gauge group,
 e.g. $A_{\alpha\dot\alpha}= A_{\alpha\dot\alpha}^a\, T^a$, with the generators satisfying 
$[T^a, T^b] = i f^{abc} T^c$ and normalized as $\tr(T^a T^b) = \frac12 \delta^{ab}$. In the special case of the $SU(2)$ gauge group
the generators are expressed in terms of Pauli matrices $T^a=\sigma^a/2$.  

As follows from  \re{La}, gauge fields, gaugino and scalars satisfy equations of motion
\begin{align}
\notag\label{eqs} 
{}& D_{\dot\alpha}^\alpha \lambda_\alpha^A - i \sqrt{2}[\phi^{AB},\bar\lambda_{\dot\alpha B}] =0
\\[1.5mm]\notag
{}& D_{\alpha\dot\alpha} \bar\lambda_A^{\dot\alpha} + i \sqrt{2} [\bar\phi_{AB},\lambda_\alpha^B] =0 
\\[1.5mm]\notag
{}& D_{\dot\alpha}^\alpha F_{\alpha\beta} + 4i  \{\lambda^A_\beta,\bar\lambda_{\dot\alpha A}\}
+[\bar\phi_{AB}, D_{\beta\dot\alpha} \phi^{AB}] =0
\\
{}&  D^2  \bar\phi_{AB} + \sqrt{2} \{\bar\lambda_{\dot\alpha A},\bar\lambda_B^{\dot\alpha}\}
- {1\over \sqrt{2}} \epsilon_{ABCD} \{\lambda^{\alpha C}, \lambda_\alpha^D\} 
+ \frac12  [\phi^{CD},[\bar\phi_{AB},\bar\phi_{CD}]]=0
\end{align}
where $D^2 \equiv D_\mu^2 = \frac12 D_{\dot\alpha}^\alpha D^{\dot\alpha}_\alpha$. The advantage of the normalization of the Lagrangian
\re{La} is that the equations of motion and their solutions do not depend on the coupling constant.

The relations \re{eqs} are invariant under (on-shell) $\mathcal N=4$ superconformal transformations. In particular,  the transformations
generated by chiral Poincare supercharges $Q_A^\alpha$ and corresponding special superconformal
generators $\bar S_A^{\dot\alpha}$ look as
\begin{eqnarray}\label{trans1}
&&
\delta A_{\alpha \dot\beta} =  -2 \zeta_\alpha^{A} \bar\lambda_{\dot\beta A}  \, , \nonumber\\
&&
\delta \phi^{AB} =  - i \sqrt{2}\zeta^{\alpha [A} \lambda_\alpha^{B]} 
\, , \nonumber\\[1mm]
&&
\delta \bar \phi_{CD} =- i \sqrt{2} \epsilon_{ABCD} \zeta^{\alpha \; A} \lambda_\alpha^B 
\, , \nonumber\\[1mm]
&&
\delta \lambda^A_\alpha =   \ft{i}2 F_{\alpha\beta} \zeta^{\beta A}  - i [\phi^{AB} , \bar\phi_{BC}] \zeta_\alpha^C
\, , \nonumber\\[1mm]
&&
\delta \bar\lambda_A^{\dot\alpha} = - \sqrt{2} \big( D^{\dot\alpha\beta} \bar\phi_{AB} \big)\zeta_\beta^B + 2 \sqrt{2}  \bar\phi_{AB}\bar\eta^{\dot\alpha B}  
\, ,
\end{eqnarray}
where brackets in the second relation denote antisymmetrization of the $SU(4)$ indices.
These relations describe transformations of fields under combined $Q-$ and $\bar S-$transformations with the corresponding parameters
being $\xi_\alpha^A$ and $\bar\eta_{\dot\alpha}^A$, respectively. 

All relations in \re{trans1} except the last one depend on the linear
($x-$dependent) combination
\begin{align}\label{xi}
\zeta_\alpha^A(x) = \xi_\alpha^A + x_{\alpha\dot\alpha}\bar\eta^{\dot\alpha A}\,.
\end{align}
This property plays an important role in our analysis and it can be understood as follows. 
We recall that $\bar S-$transformations can be realized as composition of the inversion and $Q-$transformations
\begin{align}\label{Sbar}
(\bar\eta\, \bar S) = I \cdot (\xi\, Q) \cdot I\,, \qqquad 
I(\xi_\alpha^A) = \bar\eta_{\dot\alpha}^A\,,\qqquad I(\bar\eta_{\dot\alpha}^A) = \xi_\alpha^A \,,
\end{align}
where $(\bar\eta \, \bar S) =\bar\eta^A_{\dot\alpha} \bar S_A^{\dot\alpha}$ and $(\xi\, Q)=\xi^{A\alpha}
Q_{\alpha A}$.
Let us consider the last relation in \re{trans1}. We first apply inversions and take into account that 
$I$ changes the chirality of Lorentz indices
\begin{align}\label{I-x}
I( x_{\alpha\dot\beta}) = {x_{\beta\dot\alpha}\over x^2}\,,\qqqquad
I(\bar\lambda_A^{\dot\alpha}) =  \bar\lambda_{\dot\beta A} x^{\dot\beta\alpha} x^2 
\,,\qqqquad
I(\bar\phi^{AB}) = x^2 \bar\phi^{AB}\,.
\end{align}
Then, we obtain from \re{trans1} and \re{Sbar}
\begin{align}\notag\label{QS}
(\xi\, Q)  \bar\lambda_A^{\dot\alpha} {}&= - \sqrt{2} \big( D^{\dot\alpha\beta} \bar\phi_{AB} \big)\xi_\beta^B\,,
\\
(\bar\eta\, \bar S) \bar\lambda_A^{\dot\alpha} {}&= I \cdot (\xi\, Q) \cdot I= \bar\eta^{\dot\beta B}  \, I \cdot Q_{\beta A} \cdot I (\bar\lambda_A^{\dot\alpha})
= \sqrt{2}\, \bar\eta^{B}_{\dot\beta} \, {x^{\dot\alpha}_{\beta}\over  (x^2)^2} \,I(D^{\dot\beta\beta} \bar\phi_{AB}) \,,
\end{align}
where $I(D^{\dot\beta\beta} \bar\phi_{AB}) =  x^{\dot\beta}_\gamma x^\beta_{\dot\gamma}   D^{\dot\gamma  \gamma}\,\big( x^2 \bar\phi_{AB} \big) = x^2 ( x^{\dot\beta}_\gamma x^\beta_{\dot\gamma}   D^{\dot\gamma\gamma} \, \bar\phi_{AB} - 2  x^{\dot\beta\beta}  \bar\phi_{AB})$. 

Combining the  relations \re{QS} together, we find that $\delta \bar\lambda_A^{\dot\alpha} = (\xi\cdot Q+\bar\eta\cdot \bar S) \bar\lambda_A^{\dot\alpha}$ agrees with the last relation in \re{trans1}. The additional correction to $\delta \bar\lambda$ proportional to $\bar\eta^{\dot\alpha B}$ comes from the inhomogenous term in $I(D^{\dot\beta\beta} \bar\phi_{AB})$. 
In other words, the appearance of $O(\bar\eta)$ term in the expression for $\delta \bar\lambda$ in \re{trans1} is ultimately related to the 
fact that $D^{\dot\alpha\beta} \bar\phi_{AB}$ does not transform covariantly under the inversion, or equivalently, that $\delta \bar\lambda$ 
involves the operator $D^{\dot\alpha\beta}\bar\phi^{AB}$ which is not conformal primary. 

The question arises whether the relations \re{trans1} are consistent with conformal symmetry.
Supplementing \re{trans1} with the relation 
\begin{align}\label{inv-eta}
I(\zeta_\alpha^A) =  \zeta^{\beta A}(x^{-1})_{\beta\dot\alpha} \,,
\end{align} 
that follows from \re{Sbar} and \re{I-x}, we verify that $Q+\bar S$ variations of all fields, $\delta F_{\alpha\beta}$,  $\delta F_{\dot\alpha\dot\beta}$, 
$\delta \phi^{AB}$, $\delta \lambda_\alpha^A$ and $\delta \bar\lambda^{\dot\alpha}_A$, transform under the conformal 
transformations in the same manner as the fields themselves. The same property should hold for higher order variation
of fields, e.g. for $\delta^n  \phi^{AB}$ with $n=2,3,\dots$. 

The explicit calculation of $\delta^n  \phi^{AB}$ from \re{trans1} is very cumbersome and is not efficient for higher $n$ due to proliferation of terms. Instead, we can use conformal symmetry to simply the task.
Namely, conformal symmetry restricts the possible form of $\delta^n  \phi^{AB}$ and
allows us to write its general expression in terms of a few arbitrary coefficients. The latter can be fixed by requiring the fields
to satisfy the $\mathcal N=4$ SYM equations of motion \re{eqs}.

\section{Iterative solution to the equations of motion}\label{app:B}

The equations of motion \re{eqs} are invariant under $\mathcal N=4$ superconformal transformations \re{trans1}. Following \cite{Belitsky:2000ws}, we can
exploit this property to construct solutions to \re{eqs} starting from the special solution 
\begin{align}\label{start}
F^{(0)}_{\dot\alpha\dot\beta} = \phi^{AB,(0)} = \lambda_\alpha^{A,(0)} = \bar\lambda^{\dot\alpha,(0)}_A = 0\,,
\end{align}
describing self-dual gauge field in pure Yang-Mills theory with $F^{(0)}_{\alpha\beta}\neq 0$. 
Since \re{start} verifies \re{eqs}, we can obtain 
another solution to  \re{eqs} by applying $\mathcal N=4$ superconformal transformations \re{trans1} to the fields \re{start}. 

The field configuration \re{start} is invariant under the antichiral $\bar Q + S$ transformations. Then, we use the remaining chiral $Q+\bar S$ generators to get
\begin{align}\notag\label{Phi-n}
\Phi(x;\xi,\bar\eta) 
{}& = \e^{i (\xi  Q) + i (\bar\eta  \bar S)} \Phi^{(0)}(x) \e^{-i (\xi  Q) - i (\bar\eta  \bar S)}
\\
{}& =  \Phi^{(0)} +  \Phi^{(1)} +    \Phi^{(2)} + \dots\,,
\end{align}
where $\Phi$ stands for one of the fields (scalar, gaugino and gauge field). Here $\Phi^{(0)}$ is given by \re{start} and the notation
was introduced for
\begin{align}\notag\label{Phi-n-rec}
{}& \Phi^{(1)} = i [(\xi  Q) + (\bar\eta  \bar S),\Phi^{(0)}] \equiv \delta  \Phi^{(0)}\,,
\\[2mm]
{}& \Phi^{(n)} =
{1\over n} \delta  \Phi^{(n-1)} = {1\over n!}\delta^n  \Phi^{(0)}\,,
\end{align} 
where the variation of fields $\delta\Phi$  is given by \re{trans1}. It is important to emphasise that applying \re{Phi-n-rec}
we first perform superconformal transformations \re{trans1} and, then, replace $\Phi^{(0)}$ by their explicit expressions \re{start}.

By construction, the field $\Phi(x;\xi,\bar\eta)$ depends on 
$16$ fermion modes $\xi_\alpha^A$ and $\bar\eta^{\dot\alpha A}$ (with $\alpha,\dot\alpha=1,2$ and $A=1,\dots,4$). The
second relation in \re{Phi-n} defines the expansion of $\Phi(x;\xi,\bar\eta)$ in powers of fermion modes, so that $\Phi^{(n)}$ is
a homogenous polynomial in $\xi_\alpha^A$ and $\bar\eta^{\dot\alpha A}$ of degree $n$. Its maximal degree cannot exceed
the total number of fermion modes leading to $\Phi^{(n)}=0$ for $n>16$.  
An additional condition comes from the requirement for $\Phi^{(n)}$ to have the same $R-$charge. This leads to the following
relations for different fields
\begin{align}\notag\label{dec}
{}& A_{\alpha\dot\alpha} =A_{\alpha\dot\alpha}^{(0)}+A_{\alpha\dot\alpha}^{(4)}+\dots + A_{\alpha\dot\alpha}^{(16)}\,,
\\\notag
{}& \lambda_\alpha^A = \lambda_\alpha^{A, (1)} +  \lambda_\alpha^{A, (5)}+\dots + \lambda_\alpha^{A, (13)}\,,
\\[1mm]\notag
{}& \phi^{AB} = \phi^{AB,(2)} + \phi^{AB,(6)} + \dots + \phi^{AB,(14)} \,,
\\
{}& \bar\lambda_{\dot\alpha A} =\bar\lambda_{\dot\alpha A}^{(3)} + \bar\lambda_{\dot\alpha A}^{(7)} + \dots + \bar\lambda_{\dot\alpha A}^{(15)} \,,
\end{align}
where $\bar\phi_{AB} = \frac12 \epsilon_{ABCD} \phi^{CD}$ and each subsequent term of the expansion has four fermion modes more. 

Substituting \re{dec} into \re{eqs} and matching the number
of fermion modes on both sides of the relations, we obtain the system of coupled equations for various components of fields.
To the leading order, we have from \re{eqs} 
\begin{align}\notag\label{LO}
{}&D_{\dot\alpha}^\alpha F_{\alpha\beta}^{(0)}  =    D^2 \bar\phi_{AB}^{(2)}  
- {1\over \sqrt{2}} \epsilon_{ABCD} \{\lambda^{(1),\alpha C}, \lambda_\alpha^{(1),D}\} =0\,,
\\[1.5mm]
{}& D_{\dot\alpha}^\alpha \lambda_\alpha^{(1),A}  = D_{\alpha\dot\alpha} \bar\lambda_A^{(3),\dot\alpha} + i \sqrt{2} [\bar\phi_{AB}^{(2)},\lambda_\alpha^{(1),B}] =0 \,,
\end{align}
where the covariant derivative $D_{\alpha\dot\alpha}$ is given by \re{cov} with the gauge field replaced by $A_{\alpha\dot\alpha}^{(0)}$.
To the next-to-leading order we find in a similar manner   
\begin{align}\notag\label{NLO}
{}&  D_{\dot\alpha}^\beta F^{(4)}_{\beta\alpha} 
+ [A_{\dot\alpha}^{(4),\beta}, F_{\beta\alpha}^{(0)}]  
+ 4i \{\lambda_\alpha^{(1),A},\bar\lambda^{(3)}_{\dot\alpha A}\} + [\phi^{(2),BC}, D_{\alpha\dot\alpha} \bar\phi^{(2)}_{BC}] =0\,,
\\[2mm]\notag
{}&  D^{\dot\alpha\beta} \lambda_\beta^{(5),A} + [A^{(4),\dot\alpha\beta} , \lambda_\beta^{(1),A} ] 
- i \sqrt{2}[\phi^{(2),AB},\bar\lambda_B^{(3),\dot\alpha}] = 0\,,
\\[2mm]\notag
{}&   \ft12  D^\alpha_{\dot\alpha} D_{\alpha}^{\dot\alpha}  \bar\phi_{AB}^{(6)} 
+  
  [A^{(4),\alpha}_{\dot\alpha},  D_{\alpha}^{\dot\alpha} \, \bar\phi_{AB}^{(2)} ] 
-\ft12 [ \bar\phi_{AB}^{(2)}  ,  D^\alpha_{\dot\alpha} A^{(4),\dot\alpha}_{\alpha}] + \sqrt{2} \{\bar\lambda_{\dot\alpha A}^{(3)},\bar\lambda_B^{(3),\dot\alpha}\}
 \\[2mm]
{}&      \qquad
+ \ft12 [\phi^{(2),CD},[\bar\phi_{AB}^{(2)} ,\bar\phi_{CD}^{(2)} ]] 
- \sqrt{2}  \epsilon_{ABCD} \{ \lambda^{(1),\alpha C}, \lambda_\alpha^{(5),D}\} = 0\,.
\end{align}

\subsection*{Leading order solutions}
 
Applying relations \re{Phi-n-rec} together with \re{trans1} and \re{start} we obtain the leading order corrections to scalar and gaugino
fields
\begin{align}\notag\label{LO-fields}
{}& \lambda_\alpha^{(1),A} = {i\over 2} F_{\alpha\beta}\zeta^{\beta A}\,, 
\\\notag
{}& \phi^{(2),AB} = {1\over \sqrt{2}} \zeta^{\alpha A} F_{\alpha\beta} \zeta^{\beta B} \equiv {1\over \sqrt{2}} (\zeta F \zeta)^{AB} \,, 
\\ 
{} &
\bar\lambda_{\dot\alpha A}^{(3)} 
=\frac16\epsilon_{ABCD}  \zeta^{\beta B}   (\zeta   D_{\beta\dot\alpha}  F \zeta)^{CD}  
+\epsilon_{ABCD} \bar\eta_{\dot\alpha}^B (\zeta F\zeta)^{CD}\,,
\end{align}
where $F_{\alpha\beta} = F^{(0)}_{\alpha\beta}$ is nonvanishing self-dual part of the gauge strength tensor and  $x-$dependent variable $\zeta_\alpha^A$ is defined in \re{xi}.  

It is straightforward to check that \re{LO-fields} satisfy the equations of motion \re{LO}. We verify that, in agreement with \re{Phi-n-rec}, 
the fields \re{LO-fields} are related to each other as
\begin{align}\notag\label{pat}
{}& 2\phi^{(2),AB} = \delta \phi^{(1),AB}= - i \sqrt{2}\zeta^{\alpha [A} \lambda_\alpha^{(1),B]}  \,,
\\
{}& 3\bar\lambda_A^{(3),\dot\alpha} =   \delta \bar\lambda_A^{(2),\dot\alpha}= - \sqrt{2} \big( D^{\dot\alpha\beta} \bar\phi_{AB}^{(2)} \big)\zeta_\beta^B + 2 \sqrt{2}  \bar\phi_{AB}^{(2)}\bar\eta^{\dot\alpha B} \,,
\end{align}
where expressions on the right-hand side follow from \re{trans1}.
As was mentioned above, the form of \re{LO-fields} is restricted by conformal symmetry. 
For instance, the expression for  $\bar\lambda_{\dot\alpha A}^{(3)}$ involves $(\zeta   D_{\beta\dot\alpha}  F \zeta)^{CD} = \zeta^{\gamma C}   D_{\beta\dot\alpha}  F_{\gamma\delta} \zeta ^{\delta D}   $ which is not a conformal primary operator. As was explained in Section~\ref{app:A}, this leads to the appearance of the second term in $\bar\lambda_{\dot\alpha A}^{(3)}$ proportional to $\bar\eta_{\dot\alpha}^B$ which is needed to restore correct conformal properties of the gaugino field. The relative
coefficient between the two terms in the expression for $\bar\lambda_{\dot\alpha A}^{(3)}$ is uniquely fixed by the conformal symmetry whereas the overall normalization coefficient is fixed by the
equations of motion \re{LO}.
 
\subsection*{Next-to-leading order solutions}

Direct calculation of subleading corrections to fields \re{dec} based on \re{Phi-n-rec} and \re{trans1} is very cumbersome. 
We describe here another, more efficient approach. 

We start with next-to-leading correction to the gauge field and try to construct the general expression for $A^{(4)}_{\alpha\dot\alpha}$ 
which has correct properties with respect to conformal and $R$ symmetries. By construction, $A^{(4)}_{\alpha\dot\alpha}$ is a
homogenous polynomial of degree $4$ in fermion modes $\xi^A_\alpha$ and $\bar\eta^{\dot\alpha A}$. To begin with, we look 
for an expression that depends on their linear combination $\zeta^A_\alpha$ defined in \re{xi} and has quantum numbers of 
the gauge field. Since $\zeta^A_\alpha$ has scaling dimension $(-1/2)$, the product of four $\zeta$'s should be accompanied by an 
operator carrying the scaling dimension $3$. It can only be built from the self-dual part of the strength tensor $F_{\alpha\beta}$ and
covariant derivatives $D_{\alpha\dot\alpha}$. In virtue of the equations of motion, $\epsilon^{\alpha\beta}D_{\alpha\dot\alpha}F_{\beta\gamma}=0$, such operator takes the form $D_{(\alpha\dot\alpha}F_{\beta)_\gamma}$. Contracting its Lorentz indices with those of 
the product of four $\zeta$'s we obtain $A^{(4)}_{\alpha\dot\alpha} \sim \epsilon_{ABCD} \zeta_\alpha^{A} \zeta^{\beta B} (\zeta  D_{\beta\dot\alpha} F\zeta)^{CD}$. Since this operator is not conformal primary, it should receive correction proportional to $\epsilon_{ABCD}  \zeta_\alpha^{A} \bar\eta_{\dot\alpha}^B (\zeta F\zeta)^{CD}$, the relative coefficient is fixed by the conformal symmetry. The overall normalization coefficient can be determined by requiring $A^{(4)}_{\alpha\dot\alpha}$ to satisfy the first relation in \re{NLO}. As we will show in a moment, there is
much simpler way to fix this coefficient using the second relation in \re{Phi-n-rec}.

Repeating the same analysis for
scalar and gaugino fields we get
\begin{align}\notag\label{sub-fields}
{}& A^{(4)}_{\alpha\dot\alpha}   
= - \frac1{12}\epsilon_{ABCD} \zeta_\alpha^{A} \zeta^{\beta B} (\zeta  D_{\beta\dot\alpha} F\zeta)^{CD}
- \frac12\epsilon_{ABCD}  \zeta_\alpha^{A} \bar\eta_{\dot\alpha}^B (\zeta F\zeta)^{CD}\,,
\\[2mm] \notag
{}& \lambda ^{(5),\alpha A}  = {3i\over 40}\epsilon_{CDEF} \zeta^{\alpha C} (\zeta^2)^{A D} (\zeta F^2 \zeta)^{EF}  \,,
\\
{}& \phi^{(6),AB} = - {1\over 20\sqrt{2}}\epsilon_{CDEF}   (\zeta^2)^{AC} (\zeta^2)^{BE} (\zeta F^2\zeta)^{FD}\,,
\end{align}
where the notation was introduced for
\begin{align}\notag\label{not}
{}& (\zeta^2)^{AB} = (\zeta^2)^{BA} =  \zeta^{\beta A}\epsilon_{\beta\gamma} \zeta^{\gamma B}\,,
\\ \notag
{}&(\zeta F^2 \zeta)^{EF} =-(\zeta F^2 \zeta)^{FE} =   \zeta^{\beta F} F^2_{\beta\gamma} \zeta^{\gamma E}\,,
\\
{}&F^2_{\beta\gamma} =F^2_{\gamma\beta} =F_{\beta\alpha} \epsilon^{\alpha\delta}F_{\delta\gamma}\,.
\end{align}
It is straightforward to verify that the fields \re{sub-fields} satisfy the system of coupled equations \re{NLO}.
Notice that the last two relations in \re{sub-fields} involve a conformal primary operator $F^2_{\alpha\beta}$ of dimension $4$ and,
as a consequence, the dependence on fermion modes only enters through the linear combination  $\zeta_\alpha^A$.\footnote{There is another
operator with the same quantum numbers, $D^2 F_{\alpha\beta}$, but it reduces to $-F^2_{\alpha\beta}/2$ on shell of the equations of motion.}

We recall that the subleading corrections to fields have to satisfy \re{Phi-n-rec}. In application to \re{sub-fields} these relations read
\begin{align}\notag\label{rec}
{}& 4 A^{(4)}_{\alpha\dot\alpha}  = \delta  A^{(3)}_{\alpha\dot\alpha} = -2 \zeta_\alpha^A \bar\lambda_{\dot\alpha A}^{(3)}  \,,
\\[2mm] \notag
{}& 5 \lambda^{(5),A}_\alpha =\delta \lambda^{(4),A}_\alpha=  \ft{i}2 F^{(4)}_{\alpha\beta} \zeta^{\beta A}  - i [\phi^{(2),AB} , \bar\phi_{BC}^{(2)}] \zeta_\alpha^C  \,,
\\[2mm]
{}& 6 \phi^{(6),AB} = \delta  \phi^{(5),AB}  
= -i \sqrt{2}
\zeta^{\alpha [A} \lambda ^{(5),B]}_\alpha   \,,
\end{align}
where $ F^{(4)}_{\alpha\beta} = D_{(\alpha\dot\alpha} A^{\dot\alpha,(4)}_{\beta)}$ defines the correction to self-dual part of the gauge strength
tensor. Replacing the fields with their explicit expressions \re{LO-fields} and \re{sub-fields}, we verify that the relations \re{rec} are
indeed satisfied.

We can now turn the logic around and apply the relations \re{pat} and \re{rec} to compute subleading corrections to the fields. 
Indeed, we start with the expression for $\lambda_\alpha^{(1),A}$ in \re{LO-fields} and use  \re{pat} and \re{rec} recursively to reproduce 
\re{sub-fields}. Continuing this procedure we can determine all remaining terms of the expansion \re{dec} with little efforts, e.g.
\begin{align}\label{rec1}\notag
{}& 7\bar\lambda_A^{(7),\dot\alpha} =   \delta \bar\lambda_A^{(6),\dot\alpha}= - \sqrt{2} \Big( D^{\dot\alpha\beta} \bar\phi_{AB}^{(6)}+[A^{\dot\alpha\beta,(4)} \bar\phi_{AB}^{(2)}] \Big)\zeta_\beta^B + 2 \sqrt{2}  \bar\phi_{AB}^{(6)}\bar\eta^{\dot\alpha B}\,,
\\
{}& 8 A^{(8)}_{\alpha\dot\alpha}  = \delta  A^{(7)}_{\alpha\dot\alpha} = -2 \zeta_\alpha^A \bar\lambda_{\dot\alpha A}^{(7)}\,.
\end{align}
Going through calculation of higher components of fields we find that they are proportional to the square of a fermion mode and, 
therefore, vanish 
\begin{align}
\lambda_\alpha^{(9),A} = \phi^{(10),AB} = \bar\lambda_{\dot\alpha A}^{(11)} = A_{\alpha\dot\alpha}^{(12)} =
\lambda_\alpha^{(13),A} = \phi^{(14),AB} = \bar\lambda_{\dot\alpha A}^{(15)} = A_{\alpha\dot\alpha}^{(16)} =0\,.
\end{align}
This relation implies that the expansion of fields on the instanton background \re{dec} is shorter than one might expect.
The same result was independently obtained in \cite{Bianchi}.
  
\section{Projection onto twist-two operators}\label{App:Born}

In this Appendix we explain how to use the light-ray operators \re{O-LC} to compute the correlation functions involving twist-two operators
in the Born approximation. 

To begin with, we consider the correlation function \re{G4-lc} of the light-ray operator and two half-BPS operators
in $\mathcal N=4$ SYM with the $SU(N)$ gauge group. In the Born approximation, we can
neglect gauge links in the definition \re{O-LC} of $\mathbb O(z_1,z_2)$ and express \re{G4-lc} in terms of free propagators of scalar fields $\phi^{AB}(x)=\phi^{a,AB}(x) T^a$
\begin{align}\label{free}
\vev{\phi^{a,AB}(x_1) \phi^{b,CD}(x_2)} = g^2 \delta^{ab}\epsilon^{ABCD} D(x_1-x_2)\,, 
\end{align}
where $D(x) = 1/(4\pi^2 x^2)$ and the additional factor of $g^2$ appears due to our normalization of the Lagrangian \re{La}.  

The generators of the $SU(N)$ gauge group are normalized as $\tr(T^a T^b)=\delta^{ab}/2$, so that
$ O_{\bf 20'}(x_i) = Y_{i,AB} Y_{i,CD} \phi^{a,AB}\phi^{a,CD}$.
We recall that the light-ray
operators \re{O-LC} are built out of the complex scalar field $Z=\phi^{14}$. It is convenient to represent this field as 
$Z=  Y_{Z,AB} \phi^{AB}$, with
$Y_Z$ having the only nonvanishing components $Y_{Z,14}=-Y_{Z,41}=1/2$. Then, we find
\begin{align}\notag \label{OOO-B}
{}& \vev{\mathbb O(z_1,z_2) O_{\bf 20'}(x_1)O_{\bf 20'}(x_2) }_{\rm _{Born}}  =  \frac12g^6 (N^2-1)(Y_1 Y_2) (Y_1 Y_Z) (Y_2 Y_Z)
\\[2mm]
{}&  \times  D(x_{12}) \big[ D(x_1-nz_1) D(x_2-nz_2) + D(x_1-nz_2) D(x_2-nz_1) \big]\,,
\end{align}
where $x_{12}=x_1-x_2$ and the notation was introduced for $(Y_i Y_j) =\epsilon^{ABCD} Y_{i.AB} Y_{j,CD}$. Taking into account that 
$(x_i - n z_i)^2 = x_i^2- 2 z_i (n x_i)$, we can rewrite this expression   as
\begin{align}\label{dic3}
\frac12 g^6 (N^2-1)(Y_1 Y_2) (Y_1 Y_Z) (Y_2 Y_Z){ D(x_{12}) D(x_1) D(x_2) \over (1-\epsilon_1 z_1) (1-\epsilon_2 z_2)}
 + (z_1\leftrightarrow z_2)\,,
\end{align}
with $\epsilon_i = 2(nx_i)/x_i^2$. 

To obtain the three-point correlation function of local twist-two operator, we 
substitute \re{dic3} into \re{use1}. Blowing up the integration contour in \re{use1} and picking up the
residue at $z_i=1/\epsilon_i$, we arrive at~\footnote{The powers of the coupling constant in \re{3pt-Born} and \re{2pt-Born} appear due to our definition
of the Lagrangian \re{La}. They can be removed by changing the normalization of operators $\phi^{AB} \to \phi^{AB}/g$, 
 $O_{\bf 20'} \to O_{\bf 20'}/g^2$ and  $O_S \to O_S/g^2$. }
\begin{align}\notag\label{3pt-Born}
{}& \vev{O_{S}(0)O_{\bf 20'}(1) O_{\bf 20'}(2) }_{\rm _{Born}} = \frac12 g^6 (N^2-1)  (Y_1 Y_2) (Y_1 Y_Z) (Y_2 Y_Z)
\\
{}& \qquad \times    D(x_{12}) D(x_1) D(x_2)\left[{2(nx_1)\over x_1^2} -{2(nx_2)\over x_2^2}\right]^S + (x_1\leftrightarrow x_2)\,.
\end{align}
This relation coincides with the general expression for the correlation function of twist-two operators \re{gen-exp}. Notice that 
\re{3pt-Born} vanishes for odd $S$, in agreement with the fact that the twist-two operators
carry nonnegative even spin $S$.

The same technique can be used to compute two-point correlation function of twist-two operators $\vev{O_{S}(0) \bar O_{S'}(x)}$. We start
with the correlation function of two light-ray operators separated by distance $x$. In the Born approximation we have 
\begin{align}\notag\label{OO-B}
{}& 
\vev{ \tr\big[Z(nz_1) Z(nz_2) \big] \tr\big[\bar Z(x+nz_3) \bar Z(x+nz_4) \big] }  
\\ \notag
{}& \qqqquad\qqqquad ={g^4\over 4}  (N^2-1)  D(x-nz_{13}) D(x-nz_{24}) + (z_1\leftrightarrow z_2) 
\\
 {}& \qqqquad\qqqquad
= {g^4\over 4} { (N^2-1)  D^2(x)\over  (1-\epsilon z_{13})(1-\epsilon z_{24})}  + (z_1\leftrightarrow z_2)\,,
\end{align}
with $\epsilon = 2(nx)/x^2$. Substituting this relation into \re{use1} and performing integration over $z_1$ and $z_2$, we
can project the light-ray operator $\tr\big[Z(nz_1) Z(nz_2) \big]$ onto the twist-two operator $O_S(0)$. Repeating
the same procedure with respect to $z_3$ and $z_4$, we obtain the expression for $\vev{O_{S}(0) \bar O_{S'}(x)}$ that is different from zero
only for even positive $S=S'$ and is given by
\begin{align}\label{2pt-Born}
 \vev{O_{S}(0) \bar O_{S'}(x)}_{\rm _{Born}}  = \frac12 g^4 c_{S} (N^2-1) D^2(x)\left[{2(nx)\over x^2}\right]^{2S}\delta_{SS'} \,,
\end{align}
with $c_S=(2S)!/(S!)^2$ and $D(x)$ defined in \re{free}.

\section{Instanton profile of operators}\label{App:WL}

\subsection*{Light-like Wilson line}

The calculation of the leading term $E^{(0)}(z_1,z_2)$ of the expansion of the light-like Wilson line \re{line} relies on the following identity
\begin{align}\notag\label{Sig}
i \int  dt\,  n \cdot A_{ij}^{(0)}(nt) {}& =\frac12 n^{\dot\alpha\alpha} \left[\epsilon_{i\alpha}(x_{0})_{j\dot\alpha} + \epsilon_{j\alpha}(x_{0})_{i\dot\alpha} \right] \int   
{dt\over (nt-x_0)^2 +\rho^2}
\\ {}&  
=\Sigma _{ij} \int   
{dt\, (nx_0)\over (nt-x_0)^2 +\rho^2}\,,
\end{align}
where we replaced the instanton field with its explicit expression \re{A-sym} for $x=nt -x_0$ and took into account that $n^2=0$.
Here in the second relation  we introduced the following $2\times 2$ matrix $\Sigma_{ij}$  
\begin{align}\label{sigma}
\Sigma= \Sigma_+- \Sigma_- \,,\qqquad \Sigma_+={nx_0\over 2(nx_0)} \,,\qqquad \Sigma_-= {x_0n\over 2(nx_0)}\,, 
\end{align}
with $\Sigma_\pm$ being projectors, $\Sigma_\pm^2=\Sigma_\pm$, $\Sigma_+\Sigma_-=0$ and $\Sigma_+ + \Sigma_-=1$. Notice that $\Sigma_+ n = n\, \Sigma_-= n$ and $n\,\Sigma_+=\Sigma_- n=0$.

Since the $\Sigma-$matrix in \re{Sig} does not
depend on the integration variable, the path-ordered exponential reduces to the conventional exponential leading to
\begin{align} \label{E0}
E^{(0)}(z_1,z_2) {}& = \exp\lr{\Sigma I(z_1,z_2)}  = \e^{I(z_1,z_2)} \Sigma_+  +\e^{-I(z_1,z_2)} \Sigma_- \,,
\end{align} 
where the $2\times 2$ matrices $\Sigma_\pm$ are independent on $z_i$ and $I(z_1,z_2) = - I(z_2,z_1)$ is given by
\begin{align}\label{I-fun}
  I(z_1,z_2)=\int_{z_1}^{z_2}   
{dt\, (nx_0) \over (nt-x_0)^2 +\rho^2} = -\frac12 \ln {(nz_2-x_0)^2+\rho^2 \over (nz_1-x_0)^2+\rho^2}\,.
\end{align}
We recall that the matrix indices of $\Sigma_\pm$ are identified with the $SU(2)$ indices of $E^{(0)}(z_1,z_2)$.

The first subleading correction to the Wilson line, $E^{(4)}(z_1,z_2)$, comes from the $A^{(4)}$ term in the expansion of the gauge field. It
is given by
\begin{align}
E^{(4)}(z_1,z_2) = {1\over 2}\int_{z_1}^{z_2} dt\, E^{(0)}(z_1,t) \, \bra{n}A^{(4)}(nt)|n] E^{(0)}(t,z_2)\,.
\end{align}
 Replacing $ \bra{n} A ^{(4)} |n]$ with its explicit expression \re{gauge} (evaluated for $x=nt-x_0$) and taking into account \re{E0}, we obtain
 \begin{align}\notag\label{E4}
E^{(4)} (z_1,z_2) {}&= \int_{z_1}^{z_2} dt \,  {4\rho^2\over [\rho^2+(nt-x_0)^2]^3}\epsilon_{ABCD}
 \left(\rho^2[\bar\eta^A n] - \bra{\xi^{A}}x_0|n]\right)  \vev{n\zeta^B_t}
\\
{}&\times \lr{\e^{I(z_1,t)} \Sigma_+ +\e^{-I(z_1,t)} \Sigma_-}\ket{\zeta^C_t}\bra{\zeta^D_t}  \lr{\e^{I(t,z_2)} \Sigma_+ +\e^{-I(t,z_2)} \Sigma_-}\,,
\end{align}
where $\zeta_t \equiv \zeta (tn-x_0) = \xi  + (t  n-x_0) \bar\eta$ depends on the integration variable.
Here we used shorthand notations for contraction of the indices, e.g. $(\Sigma_\pm\ket{\zeta^A})_i =(\Sigma_\pm)_i{}^j  \zeta^A_j$. 

\subsection*{Light-ray operators}

The expansion of the light-ray operator on the instanton background in powers of fermion modes takes the form
\re{OO-dec}. We show below that the last two terms of the expansion vanish, Eq.~\re{O-zero}. 
The underlying reason for this is that, by virtue of $\mathcal N=4$ superconformal symmetry, the light-ray
operator $\mathbb O(z_1,z_2)$ only depends on $12$ fermion modes, $\xi_\alpha^A$ and $[n\bar\eta^A]$.

According to its definition \re{O-LC}, the operator $\mathbb O(z_1,z_2)$ depends on scalar and gauge fields,
$Z(x)$ and $(nA(x))$, evaluated on the light-ray $x^\mu = n^\mu z$. Examining the explicit expressions for the lowest
components of these fields, Eqs.~\re{gauge} and \re{phi}, we observe that the dependence on fermion modes
enters either through $[n\bar\eta^A]$ or through the linear combination $\zeta^A_\alpha(x)$ defined in \re{xi}. For $x=n z$ the latter
simplifies as $\zeta^A_\alpha = \xi^A_\alpha + z \ket{n}_\alpha[n\bar\eta^A]$, so that the above mentioned components
of fields depend  on the light-ray on $\xi^A_\alpha$ and $[n\bar\eta^A]$ only. Then, we can use the recurrence 
relations \re{Phi-n-rec}, \re{rec} and \re{rec1} to show that the same is true for all components of fields. 
 
Thus, the light-ray operator $\mathbb O(z_1,z_2)$ depends on $12$ fermionic modes $\xi^A_\alpha$ and $[n\bar\eta^A]$
that we shall denote as $\Theta^A_i$ (with $i=1,2,3$). Then, the top component 
$\mathbb O^{(16)}(z_1,z_2)$ is necessarily proportional to the square of a fermion mode and, therefore, vanishes.
The next-to-top component $\mathbb O^{(12)}(z_1,z_2)$ contains the product of all $12$ fermion modes.
$R-$symmetry fixes its form to be 
\begin{align}\label{zero-pr}
\Theta^1 \Theta^4 \Theta^1 \Theta^4 \epsilon_{A_1B_1C_1D_1} \Theta^{A_1} \Theta^{B_1} \Theta^{C_1} \Theta^{D_1}
\epsilon_{A_2B_2C_2D_2} \Theta^{A_2} \Theta^{B_2} \Theta^{C_2} \Theta^{D_2}\,,
\end{align}
where the first four $\Theta$'s carry the $SU(4)-$charge of two scalar fields $Z=\phi^{14}$ and the remaining factors  are the
$SU(4)$ singlets. Here we did not display the lower index of $\Theta^A_i$. Counting the total number of Grassmann
variables in \re{zero-pr}, we find that it is proportional to $\Theta^1_{i_1} \Theta^1_{i_2} \Theta^1_{i_3} \Theta^1_{i_4}$.
Since the lower index can take only three values this product vanishes leading to \re{O-zero}.

\subsection*{Derivation of \re{aux3}}\label{app:der1}
 
We start by examining the instanton
profile of the product of two half-BPS operators. Using  \re{1/2BPS} we find
\begin{align}\notag\label{OO-profile}
O_{\bf 20'}^{(4)}(x_1)O_{\bf 20'}^{(4)}(x_2) = \frac14 f^2(x_1)   f^2(x_2) Y_{1,A_1B_1} Y_{1,C_1D_1}  
 Y_{2,A_2B_2} Y_{2,C_2D_2} 
\\
 \times  
 (\zeta_1^2)^{A_1C_1} (\zeta_1^2)^{B_1D_1}(\zeta_2^2)^{A_2C_2} (\zeta_2^2)^{B_2D_2}\,,
\end{align}
where $\zeta_{i}\equiv \zeta(x_i-x_0)=\xi+ (x_i-x_0) \bar\eta$ and the instanton profile function $f(x)$ is given by \re{f}
with $x\to x-x_0$.

It is convenient to rewrite the light-ray operator $\mathbb O^{(8)}(z_1,z_2)$ defined in \re{O4-8}
as $\mathbb O^{(8)}=\mathbb O^{(8)}_A+ \mathbb O^{(8)}_B + (z_1\leftrightarrow z_2)$.  
For the first term on the right-hand side we get   
\begin{align}\notag\label{1st-term}
 \mathbb O_A^{(8)}(z_1,z_2) {}&= \tr\Big[Z^{(2)}(nz_1) E^{(0)}(z_1,z_2)  Z^{(6)}(nz_2)E^{(0)}(z_2,z_1) \Big] 
\\  
{}& = {1 \over 40}  f(nz_1)f^2(nz_2) I_A^{(8)}\,,
\end{align}
where we substituted $Z^{(n)}=(Y_Z)_{AB} \phi^{(n), AB}$, replaced $\phi^{(n), AB}$ by the explicit expressions \re{phi} and introduced 
\begin{align}\label{IA}
I_A^{(8)} = (Y_Z)_{A_1 B_1}(Y_Z)_{A_2 B_2}\epsilon_{CDEF} (\zeta_{z_2}^2)^{A_2C} (\zeta_{z_2}^2)^{B_2E} 
\bra{\zeta_{z_1}^{B_1}}E^{(0)}(z_1,z_2) \ket{\zeta_{z_2}^F}\bra{\zeta_{z_2}^D}E^{(0)}(z_2,z_1)\ket{\zeta_{z_1}^{A_1}}
\end{align}
with $\zeta_{z_i}=\xi+ (z_i n-x_0) \bar\eta$. By construction,
this expression contains $8$ fermion modes.

Following \re{aux3}, we multiply expressions on the right-hand side of  \re{OO-profile} and \re{1st-term} and
integrate out $16$ fermion modes. In this way, we obtain
\begin{align}\label{inter1}
 {1\over 160}f^2(x_1)   f^2(x_2)f(nz_1)f^2(nz_2) E_{\alpha\gamma}^{(0)}(z_1,z_2)   
E_{\delta\beta}^{(0)}(z_2,z_1)   I^{\alpha\beta\gamma\delta}(Y,x)\,,
\end{align}
where the integral over fermion modes $I^{\alpha\beta\gamma\delta}$ is given by the following expression
\begin{align}\notag\label{I-f}
{}& I^{\alpha\beta\gamma\delta}(Y,x)   = (Y_Z)_{E_1 F_1}(Y_Z)_{E_2 F_2} Y_{1,A_1B_1}Y_{1,C_1D_1} Y_{2,A_2B_2}Y_{2,C_2D_2}\epsilon_{B_3C_3D_3E_3}  
\\
{}&    \times \int d^8\xi d^8\bar\eta\, (\zeta^2_1)^{A_1C_1} (\zeta^2_1)^{B_1D_1} 
  (\zeta^2_2)^{A_2C_2} (\zeta^2_2)^{B_2D_2} \zeta^{\alpha,E_1}_{z_1}\zeta^{\beta,F_1}_{z_1}  
(\zeta_{z_2}^2)^{E_2B_3} (\zeta_{z_2}^2)^{F_2C_3}\zeta_{z_2}^{\gamma,D_3}\zeta_{z_2}^{\delta,E_3}\,.
\end{align}
Taking into account that the $Y-$variables satisfy the relation $\epsilon^{ABCD} Y_{AB} Y_{CD}=0$,  we find
that the $Y-$dependence of the integral is uniquely fixed by the $SU(4)$ symmetry
\begin{align}\label{I-f1}
I^{\alpha\beta\gamma\delta} (Y,x)  = (Y_1 Y_2)(Y_1 Y_Z) (Y_2 Y_Z) \,P^{\alpha\beta\gamma\delta} (x)\,,
\end{align}
where $(Y_i Y_j) = \epsilon^{ABCD} Y_{i,AB} Y_{j,CD}$ and chiral Lorentz tensor $P^{\alpha\beta\gamma\delta} (x)$ depends
on four points, $x_{10}$, $x_{20}$, $nz_1-x_0$ and $nz_2-x_0$. The calculation of this tensor can be significantly
simplified with a help of conformal symmetry. Namely, denoting the above mentioned four points as $y_i$
(with $i=1,\dots,4$) and making use of \re{inv-eta}, we find from \re{I-f} and \re{I-f1} that  $P^{\alpha\beta\gamma\delta}$ transforms under
inversions $I(y_i^{\dot\alpha \beta}) = y_i^{\dot\beta\alpha}/y_i^2$ as
\begin{align}
I\left[P^{\alpha\beta\gamma\delta}\right] =  {y_{3}^{\dot\alpha\alpha'}y_{3}^{\dot\beta\beta'}
y_{4}^{\dot\gamma\gamma'}y_{4}^{\dot\delta\delta'}\over (y_1^2 y_2^2 y_{3}^2)^2 (y_{4}^2)^4}
P_{\alpha'\beta'\gamma'\delta'}\,.
\end{align}
This relation, combined with the condition for $f^{\alpha\beta\gamma\delta}$ to be a 
homogenous polynomial in $y_{ij}$ of degree $8$, allows us to determine $f^{\alpha\beta\gamma\delta}$ up to
an overall normalization factor. Going through calculation we find
\begin{align}\notag\label{P}
P^{\alpha\beta\gamma\delta} 
{}& = -2^4\times 3^2 \times 5\times \lr{y_{34}y_{41}y_{12}y_{24}}^{(\alpha\gamma}
\lr{y_{34}y_{41}y_{12}y_{24}}^{\beta)\delta}
\\
{}&= -2^4\times 3^2 \times 5\times z_{12}^2 \lr{n x_1x_{12}(x_2-nz_2) }^{(\alpha\gamma}
\lr{n x_1x_{12}(x_2-nz_2)}^{\beta)\delta}
\,,
\end{align}
where in the second relation we replaced $y_{ij}$ by their expressions and took into account that $n^2=0$. Here angular brackets $(\alpha\beta)$ 
denote symmetrization of Lorentz indices.  

Finally, we substitute \re{I-f1} and \re{P} into \re{inter1} and replace $F_{\alpha\beta}$ and $E^{(0)}$ with
their expressions, Eqs.~\re{F-inst} and \re{E0}, respectively. In this way, we arrive after some algebra at 
the expression that differs by the factor of $6$ from the one on the right-hand side of \re{aux3}.
\begin{align} \label{res-1st}
\int d^8\xi d^8\bar\eta\,  \mathbb O_A^{(8)}(z_1,z_2)O_{\bf 20'}^{(4)}(x_1)O_{\bf 20'}^{(4)}(x_2) 
  =  \frac16\times \text{Eq.\re{aux3}}\,.
\end{align}
For the second term on the right-hand side of \re{O4-8} we have
\begin{align} \notag\label{2nd}
  \mathbb O_B^{(8)}(z_1,z_2) {}& = \tr\Big[Z^{(2)} (nz_1) E^{(4)} (z_1,z_2) Z^{(2)} (nz_2) E^{(0)}(z_2,z_1)\Big]
    \\
{}&  =- \frac12  f(nz_1) f(nz_2)  I^{(8)}_B \,.
\end{align}
Here we replaced $Z^{(2)}=(Y_Z)_{AB} \phi^{(2), AB}$ using \re{phi} and introduced notation for 
\begin{align}
I^{(8)}_B= (Y_Z)_{A_1B_1}(Y_Z)_{A_2B_2} \bra{\zeta_{z_1}^{B_1}}
  E^{(4)}(z_1,z_2)\ket{\zeta_{z_2}^{A_2}}\bra{\zeta_{z_2}^{B_2}}E^{(0)}(z_2,z_1)\ket{\zeta_{z_1}^{A_1}}\,,
\end{align}
where  $\zeta_{z_i}=\xi+ (z_i n-x_0) \bar\eta$. Replacing $E^{(0)}$ and $E^{(4)}$ with their explicit
expressions, Eqs.~\re{E0} and \re{E4}, respectively, we get for $ I^{(8)}_B$
 \begin{align}\notag
{}&   (Y_Z)_{A_1B_1}(Y_Z)_{A_2B_2}  \epsilon_{ABCD} 
 \left(\rho^2[\bar\eta^A n] - \bra{\xi^{A}}x_0|n]\right)  \vev{n\zeta^B_0} 
\bra{\zeta^{B_2}_{z_2}} \e^{I(z_1,z_2)} \Sigma_+ +\e^{-I(z_1,z_2)} \Sigma_-\ket{\zeta_{z_1}^{A_1}}
\\
{}& \times \int_{z_1}^{z_2}   {4 \rho^2 dt\over [\rho^2+(nt-x_0)^2]^3}\bra{\zeta_{z_1}^{B_1}}  {\e^{I(z_1,t)} \Sigma_+ +\e^{-I(z_1,t)} \Sigma_-}\ket{\zeta^C_t}\bra{\zeta^D_t} {\e^{I(t,z_2)} \Sigma_+ +\e^{-I(t,z_2)} \Sigma_-}\ket{\zeta_{z_2}^{A_2}},
\end{align}
with $\zeta_{t}=\xi +(nt-x_0)\bar\eta$ and $\zeta_{0}=\xi -x_0\bar\eta$. This relation can be simplified with a help of identifies
\begin{align}
\Sigma_-\ket{\zeta_t^A} = \Sigma_-\ket{\zeta_0^A}\,,\qqqquad
\Sigma_+\ket{\zeta_t^A} = \Sigma_+\ket{\zeta_0^A} + t |\bar\eta^A]\,,
\end{align} 
that follow from \re{sigma}. Going through lengthy calculation we arrive at remarkably simple expression
\begin{align}\notag\label{I-B}
I^{(8)}_B {}& =  {8 \rho^2 (z_1-z_2)^2   \over 5(\rho^2+(nz_1-x_0)^2)(\rho^2 + (nz_2-x_0)^2)} 
\\
{}& \times  (Y_Z)_{A_1B_1}(Y_Z)_{A_2B_2}
\epsilon_{ABCD} \vev{n\zeta_0^A}\vev{n\zeta_0^B} 
  (\zeta_{z_1}^2)^{A_1C}(\zeta_{z_1}^2)^{B_1D}
[\bar\eta^{A_2} n][\bar\eta^{B_2} n]\,.
\end{align}
Then, we substitute this relation into \re{2nd}, multiply it by \re{OO-profile} and integrate over $16$ fermion modes to get
\begin{align} \label{res-2nd}
\int d^8\xi d^8\bar\eta\,  \mathbb O_B^{(8)}(z_1,z_2)O_{\bf 20'}^{(4)}(x_1)O_{\bf 20'}^{(4)}(x_2) 
  =\frac13\times \text{Eq.\re{aux3}}\,.
\end{align}
Finally, we take the sum of \re{res-1st} and \re{res-2nd}, multiply it by the factor of $2$ in order to take into account
the contribution of $(z_1\leftrightarrow z_2)$ terms  and arrive at \re{aux3}.

\subsection*{Derivation of \re{aux2}}\label{app:der2}

According to \re{shift}, the instanton profile of $\mathbb O(z_1,z_2|x)$ can be obtained from that
of $\mathbb O(z_1,z_2)$ by shifting the coordinates of all fields by $x$. As follows from \re{bare},
this transformation is equivalent to shifting the position of the instanton, $x_0\to x_0-x$. 
As before we decompose the light-ray operator as $\mathbb O^{(8)}=\mathbb O^{(8)}_A+ \mathbb O^{(8)}_B + (z_1\leftrightarrow z_2)$. Then, for the product of two operators in \re{aux2} we have 
\begin{align}\notag\label{4terms}
\mathbb O^{(8)}(z_1,z_2|0)
\bar{\mathbb O}^{(8)}(z_3,z_4|x) {}& = \mathbb O^{(8)}_A   \bar{\mathbb O}^{(8)}_A +  \mathbb O^{(8)}_B   \bar{\mathbb O}^{(8)}_B +  \mathbb O^{(8)}_A   \bar{\mathbb O}^{(8)}_B + \mathbb O^{(8)}_B   \bar{\mathbb O}^{(8)}_A  
\\[2mm]
{}& + (z_1\leftrightarrow z_2) + (z_3\leftrightarrow z_4) + (z_1\leftrightarrow z_2,z_3\leftrightarrow z_4) \,,
\end{align}
where $  \mathbb O^{(8)}_A = \mathbb O^{(8)}_A(z_1,z_2)$ and  $ \bar{\mathbb O}^{(8)}_A= \bar{\mathbb O}^{(8)}_A(z_3,z_4)\big|_{x_0\to x_0-x}$. 

Let us consider separately four terms in the first line of \re{4terms}. The instanton profile of $\mathbb O^{(8)}_A$ is given by \re{1st-term}. To get an analogous expression for $\bar{\mathbb O}^{(8)}_A$, we apply the shift $x_0\to x_0-x$ to  \re{1st-term}, change the coordinates, $z_1\to z_3$ and  $z_2\to z_4$, and replace $Y_Z$ with
conjugated $Y_{\bar Z}-$variables defined as 
 $\bar Z = \bar\phi_{14} = \phi^{23} = (Y_{\bar Z})_{AB} \phi^{AB}$ and
 satisfying $(Y_Z Y_{\bar Z})=1$. In this way we get
\begin{align}\label{O-barO-1}
 \mathbb O^{(8)}_A   \bar{\mathbb O}^{(8)}_A = {1 \over 1600}  f(nz_1)f^2(nz_2) f(x+nz_3)f^2(x+nz_4)  \times
 I_A^{(8)} \bar I_A^{(8)}\,,
\end{align}
where $\bar I_A^{(8)}$ is obtained from \re{IA} through transformations described above. Integrating out fermion
modes we find
\begin{align} \notag
\int d^8\xi d^8\bar\eta \,  I_A^{(8)} \bar I_A^{(8)} {}& = 1600\, (z_1-z_2)^2 (z_3-z_4)^2 
\bra{n} E^{(0)}(z_1,z_2) x |n]^2
\\  
{}& \times 
\Big( \bra{n} E^{(0)}(z_3,z_4) x |n]\Big|_{x_0\to x_0-x}  \Big)^2\,.
\end{align}
The matrix elements in this relation can be easily computed with a help of \re{E0} 
\begin{align}\label{E-mat}
\bra{n} E^{(0)}(z_1,z_2) x |n] = \bra{n}\lr{\e^{I(z_1,z_2)} \Sigma_+  +\e^{-I(z_1,z_2)} \Sigma_- }x |n]
= 2(xn) \e^{-I(z_1,z_2)}\,,
\end{align}  
where in the second relation we used the properties of matrices \re{sigma} and $I(z_1,z_2)$ is given by
\re{I-fun}. In this way, we obtain from \re{O-barO-1}
\begin{align}\label{tr1}
\int d^8\xi d^8\bar\eta \,  \mathbb O^{(8)}_A   \bar{\mathbb O}^{(8)}_A = {1\over 36} \times \text{Eq.\re{aux2}}\,.
\end{align}
For the second term on the right-hand side of \re{4terms} we have from \re{2nd}
\begin{align}
 \mathbb O^{(8)}_B   \bar{\mathbb O}^{(8)}_B = {1 \over 4}  f(nz_1)f(nz_2) f(x+nz_3)f(x+nz_4)  \times
 I_B^{(8)} \bar I_B^{(8)}\,,
\end{align}
where $I_B^{(8)}$ is given by \re{I-B} and $\bar I_B^{(8)}$ is obtained from $I_B^{(8)}$ through the same 
transformation as before. Integration over fermion modes yields
\begin{align}\notag
\int d^8\xi d^8\bar\eta \,  I_B^{(8)} \bar I_B^{(8)} {}& = 4096   
{[2(nx)]^4 \rho^2 (z_1-z_2)^2   \over (\rho^2+(nz_1-x_0)^2)(\rho^2 + (nz_2-x_0)^2)} 
\\
{}& \times
{\rho^2 (z_3-z_4)^2   \over (\rho^2+(x+nz_3-x_0)^2)(\rho^2 + (x+nz_4-x_0)^2)} \,,
\end{align}
leading to the following relation
\begin{align}\label{tr2}
\int d^8\xi d^8\bar\eta \,  \mathbb O^{(8)}_B   \bar{\mathbb O}^{(8)}_B ={1\over 9} \times \text{Eq.\re{aux2}}\,.
\end{align}  
For  the last two terms on the right-hand side of \re{4terms} we have from \re{1st-term} and \re{2nd}
\begin{align}\notag
 \mathbb O^{(8)}_A   \bar{\mathbb O}^{(8)}_B = - {1 \over 80}  f(nz_1)f^2(nz_2) f(x+nz_3)f(x+nz_4)  \times
 I_A^{(8)} \bar I_B^{(8)}\,,
 \\
  \mathbb O^{(8)}_B   \bar{\mathbb O}^{(8)}_A = - {1 \over 80}  f(nz_1)f(nz_2) f(x+nz_3)f^2(x+nz_4)  \times
 I_B^{(8)} \bar I_A^{(8)}\,.
\end{align}
Then, we integrate over fermion modes to get
\begin{align}
\int d^8\xi d^8\bar\eta \, I_A^{(8)} \bar I_B^{(8)} = - 2560  {[2(xn)]^2 \bra{n} E^{(0)}(z_1,z_2) x |n]^2 \rho^2 (z_3-z_4)^2   \over  (\rho^2+(nz_3-x_0)^2)(\rho^2 + (nz_4-x_0)^2)} \,,
\end{align}
The integral of $I_B^{(8)} \bar I_A^{(8)}$ is given by the same expression with variables exchanged,
$z_1\leftrightarrow z_3$ and $z_2\leftrightarrow z_4$. Using \re{E-mat} we find
\begin{align}\label{tr3}
\int d^8\xi d^8\bar\eta \,  \mathbb O^{(8)}_A   \bar{\mathbb O}^{(8)}_B =
\int d^8\xi d^8\bar\eta \,  \mathbb O^{(8)}_B   \bar{\mathbb O}^{(8)}_A = {1\over 18} \times \text{Eq.\re{aux2}}\,.
\end{align}
Combining together  \re{tr1}, \re{tr2} and \re{tr3}, we find that the sum of four terms in the first line of \re{4terms} is
$1/4  \times \text{Eq.\re{aux2}}$. Since it is invariant under the exchange of points, $z_1\leftrightarrow z_2$ and $z_3\leftrightarrow z_4$, the contribution of terms in the second line of \re{4terms} is three times larger.  As a result, the total contribution of
\re{4terms} is given by \re{aux2}.

\bibliographystyle{JHEP} 


\providecommand{\href}[2]{#2}\begingroup\raggedright\endgroup
  
\end{document}